\documentclass[times,sort&compress,3p]{elsarticle}
\journal{.}
\usepackage[labelfont=bf]{caption}
\usepackage{amsmath,amsfonts,amssymb,amsthm,booktabs,color,epsfig,graphicx,url}
\usepackage{epstopdf}
\usepackage{diagbox}
\usepackage{slashbox}

\theoremstyle{plain}
\newtheorem{theorem}{Theorem}
\newtheorem{problem}{Problem}
\newtheorem{proposition}{Proposition}
\newtheorem{lemma}{Lemma}
\newtheorem{corollary}{Corollary}
\theoremstyle{definition}

\newtheorem{remark}{Remark}
\newtheorem{example}{Example}
\begin{document}
\begin{frontmatter}

\title{Analyzing distortion riskmetrics and weighted entropy for unimodal and symmetric distributions under partial information constraints}
\author[label2,label1]{Baishuai  Zuo}
\author[label2]{Chuancun Yin\corref{mycorrespondingauthor}}
\address[label2]{School of Statistics and Data Science, Qufu Normal University, Qufu, Shandong 273165, P. R. China}
\address[label1]{Department of Mathematics, Southern University of Science and Technology, Shenzhen, Guangdong 518055, P. R. China}
\cortext[mycorrespondingauthor]{Corresponding author. Email address: \url{ccyin@qfnu.edu.cn (C. Yin)}}

\begin{abstract}
In this paper, we develop the lower and upper bounds of worst-case distortion riskmetrics and weighted entropy for unimodal, and symmetric and unimodal distributions when mean and variance informations are available. We also consider the sharp upper bounds of distortion riskmetrics and weighted entropy for symmetric distribution under known mean and variance. These results are applied to (weighted) entropies, shortfalls and other risk measures. Specifically, entropies include cumulative Tsallis past entropy, cumulative residual Tsallis entropy of order $\alpha$, extended Gini coefficient,  fractional generalized cumulative residual entropy, and fractional generalized cumulative entropy. Shortfalls include extended Gini shortfall, Gini shortfall, shortfall of cumulative residual entropy, and shortfall of cumulative residual Tsallis entropy. Other risk measures include $n$th-order expected shortfall, dual power
principle and proportional hazard
principle.
\end{abstract}

\end{frontmatter}

\section{Introduction\label{sec:1}}
A distortion riskmetric is a functional $\rho_{g} : \mathcal{X} \rightarrow \mathbb{R}$, with the following form (see Wang et al. (2020a)):
\begin{align}\label{a1}
 \rho_{g}(X)=\int_{-\infty}^{0}\left[g\left(P(X> x)\right)-g(1)\right]\mathrm{d}x+\int_{0}^{\infty}g\left(P(X> x)\right)\mathrm{d}x,
\end{align}
where  $\mathcal{X} \supset L^{\infty}$ is a law-invariant convex cone, and
$$g\in\mathcal{G}=\{ h: [0, 1] \rightarrow \mathbb{R},~ h~ \mathrm{is ~of~ bounded~ variation~ and}~h(0)=0\}.$$
The function $g$ is called the distortion function of $\rho_{g}$. Note that distortion riskmetric is not only a generalization of signed Choquet integrals (see Wang et al. (2020b)), but also a generalization of the general class of distortion risk measures (see Dhaene et al. (2012)).
Note that when $g(1)=0$ in Eq. (\ref{a1}),  $\rho_{g}(X)$ reduces to usual entropy (form) of $X$ or its distribution function $F_{X}$ given by
\begin{align}\label{a2}
 \int_{-\infty}^{+\infty}g(\overline{F}_{X}(x))\mathrm{d}x,
\end{align}
where $\overline{F}_{X}(x)= P(X> x)$.
Distortion riskmetrics have a wide range of applications in behavioral
 economics and risk management, especially in the construction of premium
principles and risk aversion (see, e.g., Yaari (1987), Denneberg (1990),  Wang (2000), Denuit et al. (2005), Dhaene et al. (2012),  Hu and Chen (2020), Pesenti et al. (2024)).

Entropy is not only a useful tool in the research of life tables and  the cost of annuities, but also plays an important role in the construction of premium
principles; see for example Haberman et al. (2011), Furman et al. (2017), Psarrakos and Sordo (2019), Hu and Chen (2020), Sun et al. (2022), Yin et al. (2023), Zuo and Yin (2023), Psarrakos et al. (2024).
For $g\in\mathcal{G}$ with $g(1)=0$, and weighted function $\psi$: $ \mathbb{R}\rightarrow \mathbb{R}$, the
 weighted entropy (form) of $X$ or $F_{X}$ based on $g$ and $\psi$ is defined by
  \begin{align}\label{a3}
 \int_{-\infty}^{+\infty}\psi(x)g(\overline{F}_{X}(x))\mathrm{d}x.
\end{align}
Eq. (\ref{a2}) is a special case of Eq. (\ref{a3}) with $\psi(x)=1$.

 In recent years, the estimation of risk measures under the conditions of model free or  distributional uncertainty has attracted widespread attention from scholars in response to the problem that many datasets in real life are model free or have distributional uncertainty.
 The monograph by R\"{u}schendorf et al. (2024) offers a state-of-the-art synthesis of research on risk bounds and model uncertainty within the area of risk management in quantitative finance and insurance. Specifically,  EI Ghaoui
et al. (2003) derived closed-form solution for the worst-case Value-at-Risk (VaR) problem under condition that the first two moments are available. Chen et al. (2011) provided closed-form solution for worst-case tail Value-at-Risk (TVaR) under condition that the first two moments are known.
Li et al. (2018) presented closed-form solution for the worst-case range Value-at-Risk (RVaR) problem when the first two moments are available. Li (2018) considered the same problem for spectral risk measures, which represent an important class of distortion risk measures that are characterized by convex distortion functions.  Zhu
and Shao (2018) generalized results of Li (2018) to derive the worst-case and best-case bounds for distortion
risk measures under additional symmetry information. Liu et al. (2020) studied the worst-case
 for a law-invariant convex risk functional when higher-order moments are known.
 Cai et al. (2025a) generalized  results of Liu et al. (2020) to consider the worst-case
 distribution for general class of distortion risk measures, and also to give explanation of such risk-averse behaviour.
  Bernard et al. (2020) provided RVaR bounds for unimodal distributions with partial
information. Bernard et al. (2025) further dealt with RVaR bounds
 for unimodal and symmetric distributions. Bernard et al. (2024) studied largest (smallest) value for the class of distortion risk measures with uncertain distributions lying in Wasserstein
ball. Cai et al. (2025b) considered the worst-case distortion risk measures of transformed losses when uncertain distributions depends on a Wasserstein
ball.
Recently,
Shao and Zhang (2023, 2024) presented
closed-form solutions of extreme-case distortion risk measures when
 the first two moments, and first and certain higher-order absolute center moments, alongside the symmetry properties of the underlying distributions, are known.
Zhao et al. (2024) dealt with the extreme-cases of a general class of distortion risk measures with  partial information. Moreover, Zuo and Yin (2025) derived worst-case distortion riskmetrics and weighted entropy for general distributions when partial information (mean and variance) is known.
But they did not consider the special characteristics of distribution, such as  unimodality and symmetry.

The motivations of this paper are as follows: First, the distortion riskmetrics and weighted entropy have a wide range of applications (see, e.g., Furman et al. (2017), Hu and Chen (2020), Wang et al. (2020a), Wang et al. (2020b), Psarrakos et al. (2024)); second, in practice, ``for a set of data, we know its partial information and special structures of distribution, such as unimodal, symmetric, and symmetric unimodal, to fast estimate a risk measure (or entropy)'' is common; third, worst-case distortion riskmetrics and weighted entropy for special structures of distribution with partial information have not been studied yet. Furthermore, Psarrakos et al. (2024) and Zuo and Yin (2025) provided the worst-case distortion riskmetrics with partial information. Neither considers a symmetric constraint. Based on this reason,  we derive the exact expressions for worst-case distortion riskmetrics and weighted entropy for symmetric distributions. For unimodal or symmetric and unimodal distributions, obtaining the exact sharp upper bound is more difficult. Inspired by Bernard et al. (2020) and Bernard et al. (2025), we attempt to obtain the upper and lower bounds of worst-case distortion riskmetrics and weighted entropy.

In this paper, we consider the following problems for distortion riskmetrics and weighted entropy defined in Eqs. (\ref{a1}) and (\ref{a3}), respectively.
\begin{problem} \label{pem.1} Consider the
following optimization problem
\begin{align*}
\sup_{X\in V (\mu,\sigma)}\rho_{g}(X),
\end{align*}
where $V (\mu,\sigma)\in \mathbb{R} \times \mathbb{R}_{+}$ takes $V_{U} (\mu,\sigma)$, $V_{S} (\mu,\sigma)$ and $V_{SU} (\mu,\sigma)$, which denote the sets of unimodal, symmetric, and  symmetric and unimodal random variables, respectively,
with mean $\mu$ and variance $\sigma^{2}$.
\end{problem}
\begin{problem}\label{pem.2} Let $\Psi$ be an increasing function with $\psi(x)=\Psi'(x)$. Consider the
following optimization problem
\begin{align*}
\sup_{X\in V^{\Psi} (\mu_{\Psi},\sigma_{\Psi})}\int_{-\infty}^{+\infty}\psi(x)g(\overline{F}_{X}(x))\mathrm{d}x,
\end{align*}
where $V ^{\Psi}(\mu_{\Psi},\sigma_{\Psi})\in \mathbb{R} \times \mathbb{R}_{+}$ takes $V _{U}^{\Psi}(\mu_{\Psi},\sigma_{\Psi})$, $V_{S}^{\Psi} (\mu_{\Psi},\sigma_{\Psi})$ and $V _{SU}^{\Psi}(\mu_{\Psi},\sigma_{\Psi})$. Here,
\begin{align*}
 V _{U}^{\Psi}(\mu_{\Psi},\sigma_{\Psi}) = \left\{X : \mathrm{E}[\Psi(F_{X}^{-1}({U}))]=\mu_{\Psi}, ~\mathrm{E}[\Psi(F_{X}^{-1}({U}))-\mu_{\Psi}]^{2}=\sigma_{\Psi}^{2}, ~\mathrm{and} ~X~is ~unimodal\right\},
\end{align*}
\begin{align*}
 V _{S}^{\Psi}(\mu_{\Psi},\sigma_{\Psi}) = \left\{X : \mathrm{E}[\Psi(F_{X}^{-1}({U}))]=\mu_{\Psi}, ~\mathrm{E}[\Psi(F_{X}^{-1}({U}))-\mu_{\Psi}]^{2}=\sigma_{\Psi}^{2}, ~\mathrm{and} ~X~is ~symmetric\right\},
\end{align*}
\begin{align*}
 V _{SU}^{\Psi}(\mu_{\Psi},\sigma_{\Psi}) = \left\{X : \mathrm{E}[\Psi(F_{X}^{-1}({U}))]=\mu_{\Psi}, ~\mathrm{E}[\Psi(F_{X}^{-1}({U}))-\mu_{\Psi}]^{2}=\sigma_{\Psi}^{2}, ~\mathrm{and} ~X~is ~symmetric ~and ~unimodal\right\},
\end{align*}
and $U$ follows a uniform distribution on $[0,1]$.
 A random variable $X$ with cdf $F_{X}$ is said to be unimodal if there exists $x_{0}\in\mathbb{R}$ (called mode) such that $F_{X}$ is convex on $(-\infty, x_{0})$ and concave on $(x_{0}, +\infty)$  (see Khintchine (1938)).
\end{problem}

The rest of this paper is organized as follows. In Section 2, we give some notations, and also recall the quantile representations of distortion riskmetrics and weighted entropy.  Section 3 contains our main results, which are the lower and upper bounds of worst-case distortion riskmetrics and weighted entropy for unimodal, symmetric, and symmetric and unimodal distributions under known mean and variance conditions.  Sections 4-6 provide some applications to (weighted) entropies, shortfalls and other risk measures, respectively. Section 7 is the numerical illustration.  Section 8 summarizes and concludes the paper.

 Throughout the paper,  assume $(\Omega, \mathcal{F}, P)$ is an atomless probability space. Let $L^{0}$
be the set of all random variables  on $(\Omega, \mathcal{F}, P)$.
Denote by $L
^{\infty}$  is
the set of essentially bounded
random variables.  We assume that the random variables have finite first two moments. Let $g'$ denote (first) right
derivative of $g$, and
  $\mathbb{I}_{A}(\cdot)$ be the indicator function of set $A$.
Furthermore,
the left-continuous generalized inverse and right-continuous generalized inverse of $F_{V}(x)=P(V\le x)$ are defined, respectively, by
$$F_{V}^{-1}(p):=\mathrm{inf}\{x\in\mathbb{R} : F_{V} (x) \geq p\},~p\in(0,1],~\mathrm{and}~F_{V}^{-1}(0):=\sup\{x\in\mathbb{R} : F_{V} (x)=0\},$$
and
$$F_{V}^{-1+}(p):=\mathrm{sup}\{x\in\mathbb{R} : F_{V} (x) \leq p\},~p\in(0,1],~\mathrm{and}~F_{V}^{-1+}(1):=F_{V}^{-1}(1).$$

\section{Preliminaries\label{sec:2}}

Quantile representations of distortion riskmetrics and weighted entropy are recalled, respectively, as follows.
\begin{lemma}\label{le.1}[Wang et al. (2020a)]
Let $g\in\mathcal{G}$, and denote $\hat{g}(u)=g(1)-g(1-u),~ u \in [0, 1]$. For $X\in L^{0}$ such that $\rho_{g}(X)$ is well defined.\\
($\mathrm{i}$) If $g$ is right-continuous, then
\begin{align*}
   \rho_{g}(X)&=\int_{0}^{1}F_{X}^{-1+}(1-u)\mathrm{d}g(u)=\int_{0}^{1}F_{X}^{-1+}(u)\mathrm{d}\hat{g}(u);
\end{align*}
($\mathrm{ii}$) If $g$ is left-continuous, then
\begin{align*}
   \rho_{g}(X)&=\int_{0}^{1}F_{X}^{-1}(1-u)\mathrm{d}g(u)=\int_{0}^{1}F_{X}^{-1}(u)\mathrm{d}\hat{g}(u);
\end{align*}
($\mathrm{iii}$) If $g$ is continuous, then
\begin{align*}
   \rho_{g}(X)&=\int_{0}^{1}F_{X}^{-1}(1-u)\mathrm{d}g(u)=\int_{0}^{1}F_{X}^{-1}(u)\mathrm{d}\hat{g}(u).
\end{align*}
\end{lemma}
\begin{lemma}\label{le.2}[Zuo and Yin (2025)]
For $g\in\mathcal{G}$ with $g(1)=0$, and denote $\hat{g}(u)=-g(1-u)$ for $u \in [0, 1]$. For a weighted function $\psi: \mathbb{R}\rightarrow \mathbb{R}$ such that $\int_{-\infty}^{+\infty}\psi(x)g(\overline{F}_{X}(x))\mathrm{d}x$ is well defined, let $\Psi$ satisfy $\psi(t)=\Psi'(t)$. \\
(i) If $g$ is right-continuous, then
\begin{align*}
   \int_{-\infty}^{+\infty}\psi(x)g(\overline{F}_{X}(x))\mathrm{d}x&=\int_{0}^{1}\Psi\left(F_{X}^{-1+}(u)\right)\mathrm{d}\hat{g}(u);
\end{align*}
(ii) If $g$ is left-continuous, then
\begin{align*}
   \int_{-\infty}^{+\infty}\psi(x)g(\overline{F}_{X}(x))\mathrm{d}x&=\int_{0}^{1}\Psi\left(F_{X}^{-1}(u)\right)\mathrm{d}\hat{g}(u);
\end{align*}
(iii) If $g$ is continuous, then
\begin{align*}
   \int_{-\infty}^{+\infty}\psi(x)g(\overline{F}_{X}(x))\mathrm{d}x&=\int_{0}^{1}\Psi\left(F_{X}^{-1}(u)\right)\mathrm{d}\hat{g}(u).
\end{align*}
\end{lemma}
Next, we give following lemmas:
\begin{lemma}\label{le.3}Let $f$ and $g$ be two functions defined on $[0,1]$ with $f(0)=g(0)$, $f(1)=g(1)$, and $g(u)\geq f(u)$ for $u\in[0,1]$. If $h(u)$ is an increasing function on $[0,1]$, and satisfies $\int_{0}^{1}h(u)\mathrm{d}f(u)<\infty$ or $\int_{0}^{1}h(u)\mathrm{d}g(u)>-\infty$, then $\int_{0}^{1}h(u)\mathrm{d}g(u)\leq\int_{0}^{1}h(u)\mathrm{d}f(u)$.
\end{lemma}
\noindent $\mathbf{Proof.}$ Using integration by parts, we can directly obtain
\begin{align*}
\int_{0}^{1}h(u)\mathrm{d}g(u)-\int_{0}^{1}h(u)\mathrm{d}f(u)=\int_{0}^{1}h(u)\mathrm{d}(g(u)-f(u))=-\int_{0}^{1}(g(u)-f(u))\mathrm{d}h(u)\leq0,
\end{align*}
where the last equality we have used that $h(u)$ is increasing function on $[0,1]$.$\hfill\square$

\begin{remark}\label{re.1}
Note that if $h(u)=F^{-1}(u)$ on $[0, 1]$, Lemma \ref{le.3} will be reduced to Lemma B.2 of Shao and Zhang (2023).
\end{remark}

As in Boyd and Vandenberghe (2004), the convex and concave envelopes for $g\in\mathcal{G}$ are defined, respectively, as
\begin{align*}
g_{\ast}&=\sup\left\{h|h:[0,1]\rightarrow \mathbb{R}~ \mathrm{is} ~\mathrm{convex}~ \mathrm{and} ~h(u)\leq g(u),~u\in[0,1]\right\},\\
g^{\ast}&=\inf\left\{h|h:[0,1]\rightarrow \mathbb{R}~ \mathrm{is}~ \mathrm{concave}~ \mathrm{and}~ h(u)\geq g(u),~u\in[0,1]\right\}.
\end{align*}
Note that $(-g)_{\ast}=-g^{\ast}$. In addition, if $g$ is convex, $g_{\ast}=g$; if $g$ is concave, $g^{\ast}=g$.

\section{Main results}
In this section, we consider the lower and upper bounds of worst-case distortion riskmetrics and  weighted entropy for unimodal, symmetric, and symmetric and unimodal distributions.

To establish formula of lower and upper bounds of worst-case distortion riskmetrics and weighted entropy for unimodal distributions, we define two sets $U_{R}^{\Psi}$ and
$U_{L}^{\Psi}$ of random variables as follows:
 \begin{align*}
 U_{R}^{\Psi} = \left\{X : \Psi(F
^{-1}
_{X} (u)) =
\begin{cases}
 a, ~\mathrm{for}~ u \in [0, b),\\
 c(u - b) + a,~ \mathrm{for} ~u \in [b, 1]
.
\end{cases}
\right\},
\end{align*}
and
  \begin{align*}
 U_{L}^{\Psi} = \left\{X : \Psi(F
^{-1}
_{X} (u)) =
\begin{cases}
 c(u - b) + a, ~\mathrm{for}~ u \in [0, b),\\
 a,~ \mathrm{for} ~u \in [b, 1]
.
\end{cases}
\right\},
\end{align*}
where $a \in \mathbb{R}$, $b \in [0, 1]$, and $c \in \mathbb{R}_{+}$.
Note that when $\Psi(x)=x$, $U_{R}^{\Psi}$ and
$U_{L}^{\Psi}$ reduce to $U_{R}$ and
$U_{L}$, respectively (see Bernard et al. (2020)):
\begin{align*}
 U_{R} = \left\{X : F
^{-1}
_{X} (u) =
\begin{cases}
 a, ~\mathrm{for}~ u \in [0, b),\\
 c(u - b) + a,~ \mathrm{for} ~u \in [b, 1]
.
\end{cases}
\right\},
\end{align*}
and
  \begin{align*}
 U_{L} = \left\{X : F
^{-1}
_{X} (u) =
\begin{cases}
 c(u - b) + a, ~\mathrm{for}~ u \in [0, b),\\
 a,~ \mathrm{for} ~u \in [b, 1]
.
\end{cases}
\right\}.
\end{align*}

\begin{theorem}\label{th.1}
Let $X\in V_{U} (\mu,\sigma)$, and $g\in\mathcal{G}$. Denote $\hat{g}(u)=g(1)-g(1-u),~ u \in [0, 1]$, and let $\hat{g}_{\ast}$ be the convex envelope  of $\hat{g}$. \\
($\mathrm{i}$) If $g$ is a concave increasing function on $[0, 1]$ and $g(1) = 1$, we have
\begin{align*}
    \sup_{X\in V_{U} (\mu,\sigma)}\rho_{g}(X)=\mu+\frac{\sigma}{3}\left[\int_{0}^{1/2}\sqrt{u(8-9u)}\mathrm{d}(\hat{g}_{\ast})'(u)+\int_{1/2}^{1}\sqrt{(9u-1)(1-u)}\mathrm{d}(\hat{g}_{\ast})'(u)\right];
\end{align*}
($\mathrm{ii}$) If $g$ is a general distortion function, then
 \begin{align*}
    \mu g(1)+\sigma\sup_{b\in[0,1]}\{\Lambda_{R}(b),\Lambda_{L}(b)\}\leq\sup_{X\in V_{U} (\mu,\sigma)}\rho_{g}(X)\leq\mu g(1)+\frac{\sigma}{3}\Theta,
\end{align*}
where
$$\Lambda_{R}(b)=\frac{-(1+b^{2})g(1)+2b\hat{g}(b)+2\int_{b}^{1}u \mathrm{d}\hat{g}(u)}{\sqrt{(1-b)^{3}(1/3+b)}},$$
$$\Lambda_{L}(b)=\sqrt{\frac{3b}{4-3b}}g(1)+\frac{2\sqrt{3}\left(-b\hat{g}(b)+\int_{0}^{b}u \mathrm{d}\hat{g}(u)\right)}{\sqrt{b^{3}(4-3b)}},$$
$$\Theta=\int_{0}^{1/2}\sqrt{u(8-9u)}\mathrm{d}(\hat{g}_{\ast})'(u)+\int_{1/2}^{1}\sqrt{(9u-1)(1-u)}\mathrm{d}(\hat{g}_{\ast})'(u).$$
\end{theorem}
\noindent $\mathbf{Proof.}$ (i) See the proof of Theorem 4.1 $(\mathrm{ii})$ in  Zhao et al. (2024).\\
 (ii) By definitions of $U_{R}$, $U_{L}$ and $V(\mu,\sigma)$, we have
\begin{align*}
 U_{R} \cap V(\mu,\sigma)= \left\{X : F
^{-1}
_{X} (u) =
\begin{cases}
 \mu-\sigma\sqrt{\frac{1-b}{1/3+b}}, ~\mathrm{for}~ u \in [0, b),\\
 \mu+\sigma\frac{2u-1-b^{2}}{\sqrt{(1-b)^{3}(1/3+b)}},~ \mathrm{for} ~u \in [b, 1]
.
\end{cases}
\right\},
\end{align*}
and
  \begin{align*}
 U_{L}\cap V(\mu,\sigma) = \left\{X : F
^{-1}
_{X} (u) =
\begin{cases}
 \mu+\sigma\frac{\sqrt{3}(2u-2b+b^{2})}{\sqrt{b^{3}(4-3b)}}, ~\mathrm{for}~ u \in [0, b),\\
 \mu+\sigma\sqrt{\frac{3b}{4-3b}},~ \mathrm{for} ~u \in [b, 1]
.
\end{cases}
\right\}.
\end{align*}
Thus,
\begin{align*}
  \nonumber \sup_{X\in  U_{R}\cap V (\mu,\sigma)}\rho_{g}(X)&=\sup_{X\in  U_{R}\cap V (\mu,\sigma)}\left[\mu g(1)+\int_{0}^{1}\left(F_{X}^{-1}(u)-\mu\right)\mathrm{d}\hat{g}(u)\right]\\
   \nonumber&=\sup_{b\in[0,1]  }\left[\mu g(1)-\sigma\int_{0}^{b}\sqrt{\frac{1-b}{1/3+b}}\mathrm{d}\hat{g}(u)+\sigma\int_{b}^{1}\frac{2u-1-b^{2}}{\sqrt{(1-b)^{3}(1/3+b)}}\mathrm{d}\hat{g}(u)\right]
   \end{align*}
   \begin{align}\label{h4}
   \nonumber&=\sup_{b\in[0,1]  }\bigg[\mu g(1)-\frac{\sigma(1+b^{2})}{\sqrt{(1-b)^{3}(1/3+b)}}g(1)+\frac{2b\sigma}{\sqrt{(1-b)^{3}(1/3+b)}}\hat{g}(b)\\
   \nonumber&~~~~+\frac{2\sigma}{\sqrt{(1-b)^{3}(1/3+b)}}\int_{b}^{1}u\mathrm{d}\hat{g}(u)\bigg]\\
   &=\mu g(1)+\sigma\sup_{b\in[0,1]  }\Lambda_{R}(b),
   \end{align}
and
\begin{align}\label{h5}
  \nonumber \sup_{X\in  U_{L}\cap V (\mu,\sigma)}\rho_{g}(X)&=\sup_{X\in  U_{L}\cap V (\mu,\sigma)}\left[\mu g(1)+\int_{0}^{1}\left(F_{X}^{-1}(u)-\mu\right)\mathrm{d}\hat{g}(u)\right]\\
   \nonumber&=\sup_{b\in[0,1]  }\left[\mu g(1)+\sigma\int_{0}^{b}\frac{\sqrt{3}(2u-2b+b^{2})}{\sqrt{b^{3}(4-3b)}}\mathrm{d}\hat{g}(u)+\sigma\int_{b}^{1}\sqrt{\frac{3b}{4-3b}}\mathrm{d}\hat{g}(u)\right]\\
   \nonumber&=\sup_{b\in[0,1]  }\bigg[\mu g(1)+\sigma\sqrt{\frac{3b}{4-3b}}g(1)-\frac{2\sqrt{3}\sigma b}{\sqrt{b^{3}(4-3b)}}\hat{g}(b)+\frac{2\sqrt{3}\sigma}{\sqrt{b^{3}(4-3b)}}\int_{0}^{b}u\mathrm{d}\hat{g}(u)\bigg]\\
   &=\mu g(1)+\sigma\sup_{b\in[0,1]}\Lambda_{L}(b).
   \end{align}
   Combining Eqs. (\ref{h4}) and (\ref{h5}), we get
   \begin{align*}
   \nonumber \sup_{X\in  (U_{L}\cup U_{R})\cap V (\mu,\sigma)}\rho_{g}(X)&=\mu g(1)+\sigma\sup_{b\in[0,1]}\{\Lambda_{L}(b),\Lambda_{R}(b)\}.
   \end{align*}
Since
$$U_{L}\cap V (\mu,\sigma)\subseteq V_{U} (\mu,\sigma),~ U_{R}\cap V(\mu,\sigma)\subseteq V_{U} (\mu,\sigma),$$
  we have
\begin{align*}
   \mu g(1)+\sigma\sup_{b\in[0,1]}\{\Lambda_{L}(b),\Lambda_{R}(b)\}= \sup_{X\in  (U_{R}\cup U_{L})\cap V (\mu,\sigma)}\rho_{g}(X)&\leq\sup_{X \in V_{U} (\mu,\sigma)}\rho_{g}(X),
   \end{align*}
   which completes the proof of the inequality on the left.

 Next, we consider the inequality on the right hand side.
  For $\hat{g}$, we can find a convex envelope $\hat{g}_{\ast}$ of $\hat{g}$ such that $\hat{g}_{\ast}\leq\hat{g}$ with $\hat{g}_{\ast}(0)=\hat{g}(0)$ and $\hat{g}_{\ast}(1)=\hat{g}(1)$.
Using Lemmas \ref{le.1} and \ref{le.3}
 we get
\begin{align}\label{h6}
   \rho_{g}(X)&=\int_{0}^{1}F_{X}^{-1+}(u)\mathrm{d}\hat{g}(u)\leq\int_{0}^{1}F_{X}^{-1+}(u)\mathrm{d}\hat{g}_{\ast}(u).
\end{align}
Since $\hat{g}_{\ast}$ is a convex function in $[0,1]$, there exists a sequence of piecewise linear functions $\tilde{g}_{1}(t) \leq \tilde{g}_{2}(t) \leq \cdots \tilde{g}_{n}(t) \leq \cdots \leq \hat{g}_{\ast}(t)$ such that $\hat{g}_{\ast}(t) = \lim_{n\rightarrow\infty} \tilde{g}_{n}(t)$, where $\tilde{g}_{n}(u)=\sum_{i=1}^{m(n)}(a_{i}^{(n)}u+b_{i}^{(n)})\mathbb{I}_{[t_{i-1}^{(n)},t_{i}^{(n)})}(u)$.
 From the monotone convergence theorem we obtain  $\lim_{n\rightarrow\infty}\int_{0}^{1}F_{X}^{-1}(u)\mathrm{d}\tilde{g}_{n}(u)=\int_{0}^{1}F_{X}^{-1}(u)\mathrm{d}\hat{g}_{\ast}(u)$.
 By the monotonicity of $\tilde{g}_{n}(X)$, combining Eq. (\ref{h6}), we have
\begin{align*}
  \sup_{X\in V_{U} (\mu,\sigma)}\rho_{g}(X)&\leq \sup_{X\in V_{U} (\mu,\sigma)}\int_{0}^{1}F_{X}^{-1+}(u)\mathrm{d}\hat{g}_{\ast}(u)\\
   &=\sup_{X\in V_{U} (\mu,\sigma)}\lim_{n\rightarrow\infty}\int_{0}^{1}F_{X}^{-1+}(u)\mathrm{d}\tilde{g}_{n}(u)\\
   &=\sup_{X\in V_{U} (\mu,\sigma)}\lim_{n\rightarrow\infty}\int_{0}^{1}F_{X}^{-1+}(u)\tilde{g}'_{n}(u)\mathrm{d}u\\
   &=\sup_{X\in V_{U} (\mu,\sigma)}\lim_{n\rightarrow\infty}\left[\mu \tilde{g}'_{n}(0)+\int_{0}^{1}(1-u)\mathrm{TVaR}_{u}(X)\mathrm{d}\tilde{g}'_{n}(u)\right]
   \end{align*}
   \begin{align}\label{zz1}
   &\nonumber\leq\lim_{n\rightarrow\infty}\left[\mu \tilde{g}'_{n}(0)+\int_{0}^{1}(1-u)\sup_{X\in V_{U} (\mu,\sigma)}\mathrm{TVaR}_{u}(X)\mathrm{d}\tilde{g}'_{n}(u)\right]\\
   &=\mu (\hat{g}_{\ast})'(0)+\int_{0}^{1}(1-u)\sup_{X\in V_{U} (\mu,\sigma)}\mathrm{TVaR}_{u}(X)\mathrm{d}(\hat{g}_{\ast})'(u).
  \end{align}

  From above sets $U_{R}\cap V(\mu,\sigma)$ and $U_{L}\cap V(\mu,\sigma)$, we can obtain $\sup_{X\in (U_{R}\cup U_{L})\cap V(\mu,\sigma)}\mathrm{TVaR}_{\alpha}(X)$. We split the
proof into two steps. $\mathbf{Step~1.}$ For  $X\in U_{R}\cap V(\mu,\sigma)$, $\mathrm{TVaR}_{\alpha}(X)$ can be expressed as
\begin{align*}
\mathrm{TVaR}_{\alpha}(X)=
\begin{cases}
\mu+\frac{\sigma(\alpha-b^{2})}{\sqrt{(1-b)^{3}(1/3+b)}},~b\in[0,\alpha),\\
\mu+\frac{\sigma\alpha}{1-\alpha}\sqrt{\frac{1-b}{1/3+b}},~b\in[\alpha,1].
\end{cases}
\end{align*}
Maximizing the expression of $\mathrm{TVaR}_{\alpha}(X)$ for $b\in[0,\alpha)$, we have following results:
For $\alpha\geq\frac{1}{3}$, the function $\mu+\frac{\sigma(\alpha-b^{2})}{\sqrt{(1-b)^{3}(1/3+b)}}$ can be maximized when $b=\frac{3\alpha-1}{2}$; for $\alpha<\frac{1}{3}$, the function $\mu+\frac{\sigma(\alpha-b^{2})}{\sqrt{(1-b)^{3}(1/3+b)}}$ can be maximized when $b=0$.
Furthermore, maximizing the expression of $\mathrm{TVaR}_{\alpha}(X)$ for $b\in[\alpha,1]$, the function $\mu+\frac{\sigma\alpha}{1-\alpha}\sqrt{\frac{1-b}{1/3+b}}$ can be maximized when $b=\alpha$.
Thus,
\begin{align}\label{yy1}
\sup_{X\in  U_{R}\cap V(\mu,\sigma)}\mathrm{TVaR}_{\alpha}(X)=
\begin{cases}
\mu+\sqrt{3}\sigma\alpha,~\alpha\in(0,\frac{1}{3}),\\
\mu+\frac{\sigma}{3}\sqrt{\frac{9\alpha-1}{1-\alpha}},~\alpha\in[\frac{1}{3},1).
\end{cases}
\end{align}
$\mathbf{Step~2.}$ For  $X\in U_{L}\cap V(\mu,\sigma)$, $\mathrm{TVaR}_{\alpha}(X)$ can be expressed as
\begin{align*}
\mathrm{TVaR}_{\alpha}(X)=
\begin{cases}
\mu+\sigma\sqrt{\frac{3b}{4-3b}},~b\in[0,\alpha],\\
\mu+\frac{\sigma\sqrt{3}\alpha}{1-\alpha}\frac{-b^{2}+2b-\alpha}{\sqrt{b^{3}(4-3b)}},~b\in(\alpha,1].
\end{cases}
\end{align*}
 Maximizing the expression of $\mathrm{TVaR}_{\alpha}(X)$ for $b\in[0,\alpha]$, the function $\mu+\sigma\sqrt{\frac{3b}{4-3b}}$ is maximized when $b=\alpha$. Moreover, maximizing the expression of $\mathrm{TVaR}_{\alpha}(X)$ for $b\in(\alpha,1]$, we have following results:
   for $\alpha>\frac{2}{3}$, the function $\mu+\frac{\sigma\sqrt{3}\alpha}{1-\alpha}\frac{-b^{2}+2b-\alpha}{\sqrt{b^{3}(4-3b)}}$ can be maximized when $b=1$; for $\alpha\leq\frac{2}{3}$, the function $\mu+\frac{\sigma\sqrt{3}\alpha}{1-\alpha}\frac{-b^{2}+2b-\alpha}{\sqrt{b^{3}(4-3b)}}$ can be maximized when $b=\frac{3\alpha}{2}$. Hence,
   \begin{align}\label{yy2}
\sup_{X\in  U_{L}\cap V(\mu,\sigma)}\mathrm{TVaR}_{\alpha}(X)=
\begin{cases}
\mu+\frac{\sigma}{3}\frac{\sqrt{\alpha(8-9\alpha)}}{1-\alpha},~\alpha\in(0,\frac{2}{3}),\\
\mu+\sqrt{3}\sigma\alpha,~\alpha\in[\frac{2}{3},1).
\end{cases}
\end{align}
Combining Eqs. (\ref{yy1}) and (\ref{yy2}), we obtain
\begin{align}\label{yy3}
\sup_{X\in (U_{R}\cup U_{L})\cap V(\mu,\sigma)}\mathrm{TVaR}_{\alpha}(X)=
\begin{cases}
\mu+\frac{\sigma}{3}\frac{\sqrt{\alpha(8-9\alpha)}}{1-\alpha},~\alpha\in(0,\frac{1}{2}),\\
\mu+\sigma\sqrt{\frac{8}{9(1-\alpha)}-1},~\alpha\in[\frac{1}{2},1).
\end{cases}
\end{align}
 Using following fact
 $$\sup_{X\in\mathcal{A}_{U}(\mu,\sigma)}\mathrm{TVaR}_{\alpha}(X)\geq\sup_{X\in V_{U}(\mu,\sigma)}\mathrm{TVaR}_{\alpha}(X)\geq \sup_{X\in (U_{R}\cup U_{L})\cap V(\mu,\sigma)}\mathrm{TVaR}_{\alpha}(X),$$
  where $\mathcal{A}_{U}(\mu,\sigma)=\left\{X : \mathrm{E}[X]=\mu, ~\mathrm{Var}[X]\leq\sigma^{2}, ~\mathrm{and} ~X~is ~unimodal\right\},$
 then combining Eq. (\ref{yy3}) and Corollary 4 of Bernard et al. (2020), we have
 \begin{align}\label{yy4}
\sup_{X\in V_{U}(\mu,\sigma)}\mathrm{TVaR}_{\alpha}(X)=
\begin{cases}
\mu+\frac{\sigma}{3}\frac{\sqrt{\alpha(8-9\alpha)}}{1-\alpha},~\alpha\in(0,\frac{1}{2}),\\
\mu+\sigma\sqrt{\frac{8}{9(1-\alpha)}-1},~\alpha\in[\frac{1}{2},1).
\end{cases}
\end{align}
 Substituting Eq. (\ref{yy4}) into (\ref{zz1}) we instantly obtain the desired results.$\hfill\square$
\begin{remark}\label{re.2}
Note that if $g$ is an increasing function on $[0, 1]$ with $g(1) = 1$, Theorem \ref{th.1} (ii) reduces to Theorem 4.1 $(\mathrm{iii})$ of  Zhao et al. (2024).
\end{remark}

 \begin{corollary}\label{co.1}
 Let $g\in\mathcal{G}$ with $g(1) = 0$. Denote $\hat{g}(u)=-g(1-u)$, $u\in[0,1]$, and let $\hat{g}_{\ast}$ be a convex envelope  of $\hat{g}$. Then
\begin{align*}
  \sigma\sup_{b\in[0,1]}\{\Lambda_{R}(b),\Lambda_{L}(b)\}\leq \sup_{X\in V_{U} (\mu,\sigma)}\int_{-\infty}^{+\infty}g(\overline{F}_{X}(x))\mathrm{d}x\leq\frac{\sigma}{3}\Theta,
\end{align*}
where
$$\Lambda_{R}(b)=\frac{2b\hat{g}(b)+2\int_{b}^{1}u \mathrm{d}\hat{g}(u)}{\sqrt{(1-b)^{3}(1/3+b)}},
~~\Lambda_{L}(b)=\frac{2\sqrt{3}\left(-b\hat{g}(b)+\int_{0}^{b}u \mathrm{d}\hat{g}(u)\right)}{\sqrt{b^{3}(4-3b)}},$$
$$\Theta=\int_{0}^{1/2}\sqrt{u(8-9u)}\mathrm{d}(\hat{g}_{\ast})'(u)+\int_{1/2}^{1}\sqrt{(9u-1)(1-u)}\mathrm{d}(\hat{g}_{\ast})'(u).$$

\end{corollary}

For convenience, we denote $k_{\ast}'$ as first taking the convex envelope of $k$ and then taking its derivative.

\begin{theorem}\label{th.2} Let $g\in\mathcal{G}$, and $g$ be continuous, then
 \begin{align}\label{a7}
   \sup_{X\in V_{S} (\mu,\sigma)}\rho_{g}(X)=\mu g(1)+
   \frac{\sigma}{2}\sqrt{\int_{0}^{1}[(\hat{g}(u)+\hat{g}(1-u))_{\ast}']^{2}\mathrm{d}u},
\end{align}
where $\hat{g}(u)=g(1)-g(1-u),~ u \in [0, 1]$.
If $(\hat{g}(u)+\hat{g}(1-u))_{\ast}'=0$ is almost everywhere (a.e.), then (\ref{a7}) can be obtained by any random variable $X \in
V_{S} (\mu, \sigma)$; If $(\hat{g}(u)+\hat{g}(1-u))_{\ast}'\neq 0 ~(a.e.)$, then (\ref{a7})
 can be obtained by the
worst-case distribution of random variable $X_{\ast}$ characterized by
$$F_{X_{\ast}}^{-1}(u)=\mu+\sigma\frac{(\hat{g}(u)+\hat{g}(1-u))_{\ast}'}{\sqrt{\int_{0}^{1}\left((\hat{g}(v)+\hat{g}(1-v))_{\ast}'\right)^{2}\mathrm{d}v}}.$$
\end{theorem}
\noindent $\mathbf{Proof.}$ For $X\in V_{S} (\mu,\sigma)$, it follows that $F_{X}^{-1}(u)+F_{X}^{-1}(1-u)=2\mu~(a.e.)$. Then $\rho_{g}(X)$ can be written as
\begin{align*}
\rho_{g}(X)&=\frac{1}{2}\left[\int_{0}^{1}F_{X}^{-1}(1-u)\mathrm{d}g(u)+\int_{0}^{1}F_{X}^{-1}(u)\mathrm{d}\hat{g}(u)\right]\\
&=\mu g(1)+\frac{1}{2}\int_{0}^{1}F_{X}^{-1}(u)\mathrm{d}(\hat{g}(u)+\hat{g}(1-u)).
\end{align*}
Using same arguments as those in Theorem 3.1 of Zuo and Yin (2025), we can obtain Eq. (\ref{a7}). $\hfill\square$

For the general $g$, to derive worst-case for symmetric distribution, we define $\check{g}(x)=\max\{g(x-),g(x+),~g(x)\}$ for $x\in(0,1)$, and $\check{g}(x)=g(x)$ for $x=0 ~and ~1$.
 \begin{proposition}\label{prop.1} Let $g\in\mathcal{G}$.  Then
 \begin{align}\label{i}
   \sup_{X\in V_{S} (\mu,\sigma)}\rho_{g}(X)=
   \mu g(1)+\sigma\sqrt{\int_{0}^{1}[(\bar{g}^{\ast})'(1-u)]^{2}\mathrm{d}u},
\end{align}
where $\bar{g}(u)=\frac{g(u)+g(1-u)-g(1)}{2},~ u \in [0, 1]$. Furthermore, if $g=\check{g}$,
and $(\bar{g}^{\ast})'(u)=0$ is almost everywhere (a.e.), then (\ref{i}) can be obtained by any random variable $X \in
V_{S} (\mu, \sigma)$;
 If $g=\check{g}$, and $(\bar{g}^{\ast})'(u)\neq 0 ~(a.e.)$, then (\ref{i})
 can be obtained by the
worst-case distribution of random variable $X_{\ast}$ characterized by
$$F_{X_{\ast}}^{-1}(u)=\mu+\sigma\frac{(\bar{g}^{\ast})'(1-u)}{\sqrt{\int_{0}^{1}\left((\bar{g}^{\ast})'(1-v)\right)^{2}\mathrm{d}v}}.$$
 \end{proposition}
 \noindent $\mathbf{Proof.}$ For $X\in V_{S} (\mu,\sigma)$, using definition of $\rho_{g}(X)$ and  Proposition 2 (iii) of Wang et al. (2020a) we can obtain
 \begin{align}\label{iii}
 \rho_{g}(X)=\mu g(1)+\rho_{\bar{g}}(X).
 \end{align}
 From Theorem 5 of Pesenti et al. (2024), we have
 \begin{align}\label{ii}
 \sup_{X\in V (\mu,\sigma)}\rho_{\bar{g}}(X)=\sigma\sqrt{\int_{0}^{1}[(\bar{g}^{\ast})'(1-u)]^{2}\mathrm{d}u}.
  \end{align}
  Moreover, if $\bar{g}=\check{\bar{g}}$, and $(\bar{g}^{\ast})'(u)=0$ is almost everywhere (a.e.), then (\ref{ii}) can be obtained by any random variable $X \in
V_{S} (\mu, \sigma)$;
 If $\bar{g}=\check{\bar{g}}$, and $(\bar{g}^{\ast})'(u)\neq 0 ~(a.e.)$, then (\ref{ii})
 can be obtained by the
worst-case distribution of random variable $X_{\ast}$ characterized by
$$F_{X_{\ast}}^{-1}(u)=\mu+\sigma\frac{(\bar{g}^{\ast})'(1-u)}{\sqrt{\int_{0}^{1}\left((\bar{g}^{\ast})'(1-v)\right)^{2}\mathrm{d}v}}.$$
Note that $\sup_{X\in V (\mu,\sigma)}\rho_{\bar{g}}(X)\geq\sup_{X\in V_{S} (\mu,\sigma)}\rho_{\bar{g}}(X)$ and $X\in V_{S} (\mu,\sigma)$. Thus, \begin{align}\label{iv}
\sup_{X\in V_{S} (\mu,\sigma)}\rho_{\bar{g}}(X)=\sup_{X\in V (\mu,\sigma)}\rho_{\bar{g}}(X).
\end{align}
Combining Eqs. (\ref{iii}), (\ref{ii}) and (\ref{iv}), we obtain Eq. (\ref{i}). $\hfill\square$

\begin{remark}\label{re.3}
Note that if $g=\check{g}$, and $g$ is an increasing function on $[0, 1]$ with $g(1) = 1$, Proposition \ref{prop.1} reduces to Proposition 4.2 $(\mathrm{i})$ of  Zhao et al. (2024).
\end{remark}

\begin{corollary}\label{co.2}
 For $g\in\mathcal{G}$ and continuous $g$ with $g(1) = 0$, the sharp upper bound of entropy  for symmetric distributions is given by
 \begin{align}\label{a8}
   \sup_{X\in V_{S} (\mu,\sigma)}\int_{-\infty}^{+\infty}g(\overline{F}_{X}(x))\mathrm{d}x= \frac{\sigma}{2}\sqrt{\int_{0}^{1}[(\hat{g}(u)+\hat{g}(1-u))_{\ast}']^{2}\mathrm{d}u},
\end{align}
where $\hat{g}(u)=-g(1-u),~ u \in [0, 1]$.
If $(\hat{g}(u)+\hat{g}(1-u))_{\ast}'=0$ is almost everywhere (a.e.), then (\ref{a8}) can be obtained by any random variable $X \in
V _{S}(\mu, \sigma)$; If $(\hat{g}(u)+\hat{g}(1-u))_{\ast}'\neq0 ~(a.e.)$, then (\ref{a8})
 can be obtained by the
worst-case distribution of random variable $X_{\ast}$ characterized by
$$F_{X_{\ast}}^{-1}(u)=\mu+\sigma\frac{(\hat{g}(u)+\hat{g}(1-u))_{\ast}'}{\sqrt{\int_{0}^{1}\left((\hat{g}(v)+\hat{g}(1-v))_{\ast}'\right)^{2}\mathrm{d}v}}.$$
\end{corollary}
To derive formula of lower and upper bounds of worst-case distortion riskmetrics for symmetric unimodal distribution, we define set $S(\alpha)$ for $\alpha\in[0,1]$ as follows (see Bernard et al. (2025)):
\begin{align*}
S(\alpha)=\left\{X:~F_{Y}^{-1}(u)=
\begin{cases}
a+c(u-1+b),~\mathrm{for}~u\in(0,1-b),\\
a,~\mathrm{for}~u\in[1-b,b],\\
a+c(u-b),~\mathrm{for}~u\in(b,1),
\end{cases}
where~(a,b,c)\in \mathbb{R}\times [\frac{1}{2},\max(\alpha,1-\alpha)]\times\mathbb{R}^{+}.
\right\}
\end{align*}
\begin{theorem}\label{th.3}Let  $g\in\mathcal{G}$, $\hat{g}(u)=g(1)-g(1-u),~ u \in [0, 1]$, and $\hat{g}_{\ast}$ be a convex envelope  of $\hat{g}$.\\
($\mathrm{i}$) If $g$ is a concave increasing function on $[0, 1]$ and $g(1) = 1$, we have
 \begin{align*}
   \sup_{X\in V_{SU} (\mu,\sigma)}\rho_{g}(X)= \mu +\sigma\left[\frac{2}{3}\int_{0}^{\frac{1}{3}}\sqrt{u}\mathrm{d}\hat{g}_{\ast}'(u)+\sqrt{3}\int_{\frac{1}{3}}^{\frac{2}{3}}(1-u)u\mathrm{d} \hat{g}_{\ast}'(u)+\frac{2}{3}\int_{\frac{2}{3}}^{1}\sqrt{1-u}\mathrm{d}\hat{g}_{\ast}'(u)\right];
\end{align*}
$(\mathrm{ii})$ If $g$ is a general distortion function, then
 \begin{align*}
   \mu g(1)+\sigma\sup_{b\in[\frac{1}{2},1)}\Upsilon(b)\leq \sup_{X\in V_{SU} (\mu,\sigma)}\rho_{g}(X)\leq \mu g(1)+\sigma\Delta,
\end{align*}
where
$$\Upsilon(b)=\frac{(b-1)\hat{g}(1-b)-b (g(1)-\hat{g}(b))+\int_{0}^{1-b}u \mathrm{d}\hat{g}(u)+\int_{b}^{1}u\mathrm{d}\hat{g}(u)}{\sqrt{2/3(1-b)^{3}}},$$
$$\Delta=\frac{2}{3}\int_{0}^{\frac{1}{3}}\sqrt{u}\mathrm{d}\hat{g}_{\ast}'(u)+\sqrt{3}\int_{\frac{1}{3}}^{\frac{2}{3}}(1-u)u\mathrm{d}\hat{g}_{\ast}'(u)+\frac{2}{3}\int_{\frac{2}{3}}^{1}\sqrt{1-u}\mathrm{d}\hat{g}_{\ast}'(u).$$

\end{theorem}
\noindent $\mathbf{Proof.}$ (i) See the proof of Theorem 4.3 $(\mathrm{ii})$ in  Zhao et al. (2024).\\
 (ii) By definitions of $S(\alpha)$ and $V(\mu,\sigma)$, we have
\begin{align*}
S(\alpha)\cap V(\mu,\sigma)=\left\{X:~F_{X}^{-1}(u)=
\begin{cases}
\mu+\sigma\frac{u-1+b}{\sqrt{2/3(1-b)^3}},~\mathrm{for}~u\in(0,1-b),\\
\mu,~\mathrm{for}~u\in[1-b,b],\\
\mu+\sigma\frac{u-b}{\sqrt{2/3(1-b)^{3}}},~\mathrm{for}~u\in(b,1),
\end{cases}~ \mathrm{where}~ b\in[\frac{1}{2}, \max(\alpha, 1 - \alpha)],~\alpha\in[0,1]\right\}.
\end{align*}
 Thus,
\begin{align*}
  \nonumber \sup_{\alpha\in[0,1]}\sup_{X\in S(\alpha)\cap V(\mu,\sigma)}\rho_{g}(X)&=\sup_{\alpha\in[0,1]}\sup_{X\in S(\alpha)\cap V(\mu,\sigma)}\left[\mu g(1)+\int_{0}^{1}\left(F_{X}^{-1}(u)-\mu\right)\mathrm{d}\hat{g}(u)\right]\\
   \nonumber&=\sup_{b\in [\frac{1}{2},1)}\left[\mu g(1)+\sigma\int_{0}^{1-b}\frac{u-1+b}{\sqrt{2/3(1-b)^{3}}}\mathrm{d}\hat{g}(u)+\sigma\int_{b}^{1}\frac{u-b}{\sqrt{2/3(1-b)^{3}}}\mathrm{d}\hat{g}(u)\right]\\
   \nonumber&=\sup_{b\in [\frac{1}{2},1)}\left\{\mu g(1)+\frac{\sigma\left[(b-1)\hat{g}(1-b)-b (g(1)-\hat{g}(b))+\int_{0}^{1-b}u \mathrm{d}\hat{g}(u)+\int_{b}^{1}u\mathrm{d}\hat{g}(u)\right]}{\sqrt{2/3(1-b)^{3}}}\right\}\\
   &=\mu g(1)+\sigma\sup_{b\in [\frac{1}{2},1)}\Upsilon(b).
   \end{align*}
    Since
$$S(\alpha)\cap V(\mu,\sigma)\subseteq V_{SU} (\mu,\sigma),$$
  we have
   $$\mu g(1)+\sigma\sup_{b\in[\frac{1}{2},1)}\Upsilon(b)= \sup_{X\in S(\alpha)\cap V (\mu,\sigma)}\rho_{g}(X)\leq \sup_{X \in V_{SU} (\mu,\sigma)}\rho_{g}(X).$$

  Next, we consider the inequality on the other side. From above set $S(\alpha)\cap V(\mu,\sigma)$, we can obtain\\ $\sup_{X\in S(\alpha)\cap V(\mu,\sigma)}\mathrm{TVaR}
_{\alpha}(X)$. The proof can then be divided into two cases: $\alpha\geq\frac{1}{2}$ and $\alpha<\frac{1}{2}$. \\
$\mathbf{Case ~1:}$ When $\alpha\geq\frac{1}{2}$, for $X\in S(\alpha)\cap V(\mu,\sigma)$, $\mathrm{TVaR}_{\alpha}(X)$ can be expressed as $\mathrm{TVaR}_{\alpha}(X)=\mu+\frac{\sigma}{2(1-b)}\sqrt{\frac{3}{2(1-b)}}(\alpha+1-2b)$. Maximizing expression of $\mathrm{TVaR}_{\alpha}(X)$ for $b\in[\frac{1}{2},\alpha]$. We have following results: For $\alpha\geq\frac{2}{3}$, the function $\mu+\frac{\sigma}{2(1-b)}\sqrt{\frac{3}{2(1-b)}}(\alpha+1-2b)$ can be maximized when $b=\frac{3\alpha-1}{2}$; for $\alpha<\frac{2}{3}$, the function $\mu+\frac{\sigma}{2(1-b)}\sqrt{\frac{3}{2(1-b)}}(\alpha+1-2b)$ can be maximized when $b=\frac{1}{2}$. Thus,
\begin{align*}
  \sup_{X\in S(\alpha)\cap V(\mu,\sigma)}\mathrm{TVaR}
_{\alpha}(X)=\begin{cases}
 \mu +\sigma\sqrt{3}\alpha,~\alpha\in[\frac{1}{2},\frac{2}{3}),\\
\mu +\sigma\frac{2}{3\sqrt{1-\alpha}},~\alpha\in[\frac{2}{3},1).
\end{cases}
\end{align*}
$\mathbf{Case ~2:}$ When $\alpha<\frac{1}{2}$, for $X\in S(\alpha)\cap V(\mu,\sigma)$, $\mathrm{TVaR}_{\alpha}(X)$ can be expressed as\\ $\mathrm{TVaR}_{\alpha}(X)=\mu+\frac{\alpha}{1-\alpha}\frac{\sigma}{2(1-b)}\sqrt{\frac{3}{2(1-b)}}(2-2b-\alpha)$.
 Maximizing expression of $\mathrm{TVaR}_{\alpha}(X)$ for $b\in[\frac{1}{2},1-\alpha]$. Then we have following results: For $\alpha<\frac{1}{3}$, the function $\mu+\frac{\alpha}{1-\alpha}\frac{\sigma}{2(1-b)}\sqrt{\frac{3}{2(1-b)}}(2-2b-\alpha)$ can be maximized when $b=1-\frac{3\alpha}{2}$; for $\alpha\geq\frac{1}{3}$, the function $\mu+\frac{\alpha}{1-\alpha}\frac{\sigma}{2(1-b)}\sqrt{\frac{3}{2(1-b)}}(2-2b-\alpha)$ can be maximized when $b=\frac{1}{2}$. Hence,
 \begin{align*}
  \sup_{X\in S(\alpha)\cap V(\mu,\sigma)}\mathrm{TVaR}
_{\alpha}(X)=\begin{cases}
\mu +\sigma\frac{2\sqrt{\alpha}}{3(1-\alpha)},~\alpha\in(0,\frac{1}{3})\\
\mu +\sigma\sqrt{3}\alpha,~\alpha\in[\frac{1}{3},\frac{1}{2}).
\end{cases}
\end{align*}
 Combining $\mathbf{Cases ~1}$ and $\mathbf{2}$ we have
 \begin{align}\label{xx1}
\sup_{X\in S(\alpha)\cap V(\mu,\sigma)}\mathrm{TVaR}
_{\alpha}(X)=
 \begin{cases}
 \mu +\sigma\frac{2\sqrt{\alpha}}{3(1-\alpha)},~\alpha\in(0,\frac{1}{3}),\\
 \mu +\sigma\sqrt{3}\alpha,~\alpha\in[\frac{1}{3},\frac{2}{3}),\\
\mu +\sigma\frac{2}{3\sqrt{1-\alpha}},~\alpha\in[\frac{2}{3},1).
\end{cases}
\end{align}
 Let $\beta \rightarrow 1$ in Theorem 4.1 of Bernard et al. (2025), we obtain
 \begin{align}\label{bd}
\sup_{X\in \mathcal{A}_{SU} (\mu,\sigma)}\mathrm{TVaR}
_{\alpha}(X)=
 \begin{cases}
 \mu +\sigma\frac{2\sqrt{\alpha}}{3(1-\alpha)},~\alpha\in(0,\frac{1}{3}),\\
 \mu +\sigma\sqrt{3}\alpha,~\alpha\in[\frac{1}{3},\frac{2}{3}),\\
\mu +\sigma\frac{2}{3\sqrt{1-\alpha}},~\alpha\in[\frac{2}{3},1),
\end{cases}
\end{align}
where $\mathcal{A}_{SU} (\mu,\sigma)=\left\{X : \mathrm{E}[X]=\mu, ~\mathrm{Var}[X]\leq\sigma^{2}, ~\mathrm{and} ~X~is ~symmetric ~and ~unimodal\right\}.$
Using the fact
 $$\sup_{X\in \mathcal{A}_{SU} (\mu,\sigma)}\mathrm{TVaR}
_{\alpha}(X)\geq \sup_{X\in V_{SU} (\mu,\sigma)}\mathrm{TVaR}
_{\alpha}(X)\geq\sup_{X\in S(\alpha)\cap V(\mu,\sigma)}\mathrm{TVaR}
_{\alpha}(X),$$
then combining Eqs.  (\ref{xx1}) and (\ref{bd}) we obtain
\begin{align}\label{xx2}
\sup_{X\in V_{SU} (\mu,\sigma)}\mathrm{TVaR}
_{p}(X)=\begin{cases}
 \mu +\sigma\frac{2\sqrt{p}}{3(1-p)},~p\in(0,\frac{1}{3}),\\
 \mu +\sigma\sqrt{3}p,~p\in[\frac{1}{3},\frac{2}{3}),\\
\mu +\sigma\frac{2}{3\sqrt{1-p}},~p\in[\frac{2}{3},1).
\end{cases}
\end{align}
Similar to the proof of Theorem \ref{th.1}, using Eq. (\ref{xx2}),
 we can obtain the required results.
$\hfill\square$
\begin{remark}\label{re.4}
Note that if $g$ is an increasing function on $[0, 1]$ with $g(1) = 1$, Theorem \ref{th.3} (ii) reduces to Theorem 4.3 $(\mathrm{iii})$ of  Zhao et al. (2024).
\end{remark}
\begin{corollary}\label{co.3}
 Let $g\in\mathcal{G}$ with $g(1) = 0$. Denote $\hat{g}(u)=-g(1-u),~ u \in [0, 1]$, and let $\hat{g}_{\ast}$ is a convex envelope  of $\hat{g}$.  Then
\begin{align*}
   \sigma\sup_{b\in[\frac{1}{2},1)}\Upsilon(b)\leq\sup_{X\in V_{SU} (\mu,\sigma)}\int_{-\infty}^{+\infty}g(\overline{F}_{X}(x))\mathrm{d}x\leq\sigma\Delta,
\end{align*}
where
$$\Upsilon(b)=\frac{(b-1)\hat{g}(1-b)+b \hat{g}(b)+\int_{0}^{1-b}u \mathrm{d}\hat{g}(u)+\int_{b}^{1}u\mathrm{d}\hat{g}(u)}{\sqrt{2/3(1-b)^{3}}},$$
$$\Delta=\frac{2}{3}\int_{0}^{\frac{1}{3}}\sqrt{u}\mathrm{d}\hat{g}_{\ast}'(u)+\sqrt{3}\int_{\frac{1}{3}}^{\frac{2}{3}}(1-u)u\mathrm{d}\hat{g}_{\ast}'(u)+\frac{2}{3}\int_{\frac{2}{3}}^{1}\sqrt{1-u}\mathrm{d}\hat{g}_{\ast}'(u).$$
\end{corollary}

\begin{theorem}\label{th.4}Let $g\in\mathcal{G}$ with $g(1)=0$, and let $\Psi$ be an increasing function such that $\Psi'(x)=\psi(x)$ is the weighted function. Then,
 \begin{align*}
   \sigma_{\Psi}\sup_{b\in[0,1]}\{\Lambda_{R}(b),\Lambda_{L}(b)\}&\leq\sup_{X\in V_{U}^{\Psi} (\mu_{\Psi},\sigma_{\Psi})}\int_{-\infty}^{+\infty}\psi(x)g(\overline{F}_{X}(x))\mathrm{d}x\leq\frac{\sigma_{\Psi}}{3}\Theta,
\end{align*}
where
$$\Lambda_{R}(b)=\frac{2b\hat{g}(b)+2\int_{b}^{1}u \mathrm{d}\hat{g}(u)}{\sqrt{(1-b)^{3}(1/3+b)}},
~~\Lambda_{L}(b)=\frac{2\sqrt{3}\left(-b\hat{g}(b)+\int_{0}^{b}u \mathrm{d}\hat{g}(u)\right)}{\sqrt{b^{3}(4-3b)}},$$
$$\Theta=\int_{0}^{1/2}\sqrt{u(8-9u)}\mathrm{d}\hat{g}_{\ast}'(u)+\int_{1/2}^{1}\sqrt{(9u-1)(1-u)}\mathrm{d}\hat{g}_{\ast}'(u).$$ $\hat{g}(u)=-g(1-u)$, $u\in[0,1]$, and $\hat{g}_{\ast}$ is a convex increasing envelope  of $\hat{g}$.
\end{theorem}
\noindent  $\mathbf{Proof.}$ By definitions of $U_{R}^{\Psi}$, $U_{L}^{\Psi}$ and $V^{\Psi}(\mu_{\Psi},\sigma_{\Psi})$, we have
\begin{align*}
U_{R}^{\Psi}\cap V^{\Psi}(\mu_{\Psi},\sigma_{\Psi})=\left\{X:~\Psi(F_{X}^{-1}(u))=
\begin{cases}
\mu_{\Psi}-\sigma_{\Psi}\sqrt{\frac{1-b}{1/3+b}},~\mathrm{for}~u\in[0,b),\\
\mu_{\Psi}+\sigma_{\Psi}\frac{2u-1-b^{2}}{\sqrt{(1-b)^{3}(1/3+b)}},~\mathrm{for}~u\in[b,1],
\end{cases}~b\in[0,1]\right\}
\end{align*}
and
\begin{align*}
U_{L}^{\Psi}\cap V^{\Psi}(\mu_{\Psi},\sigma_{\Psi})=\left\{X:~\Psi(F_{X}^{-1}(u))=
\begin{cases}
\mu_{\Psi}+\sigma_{\Psi}\frac{\sqrt{3}(2u-2b+b^{2})}{\sqrt{b^{3}(4-3b)}},~\mathrm{for}~u\in[0,b),\\
\mu_{\Psi}+\sigma_{\Psi}\sqrt{\frac{3b}{4-3b}},~\mathrm{for}~u\in[b,1].
\end{cases}~b\in[0,1]\right\}.
\end{align*}
Thus,
\begin{align}\label{h9}
   \nonumber\sup_{X\in U_{R}^{\Psi}\cap V^{\Psi}(\mu_{\Psi},\sigma_{\Psi})}\int_{-\infty}^{+\infty}\psi(x)g(\overline{F}_{X}(x))\mathrm{d}x &=\sup_{X\in U_{R}^{\Psi}\cap V^{\Psi}(\mu_{\Psi},\sigma_{\Psi})}\int_{0}^{1}\left(\Psi\left(F_{X}^{-1}(u)\right)-\mu_{\Psi}\right)\mathrm{d}\hat{g}(u)\\
   \nonumber&=\sup_{b\in [0,1]}\left[-\sigma_{\Psi}\int_{0}^{b}\sqrt{\frac{1-b}{1/3+b}}\mathrm{d}\hat{g}(u)+\sigma_{\Psi}\int_{b}^{1}\frac{2u-1-b^{2}}{\sqrt{(1-b)^{3}(1/3+b)}}\mathrm{d}\hat{g}(u)\right]\\
   &=\sigma_{\Psi}\sup_{b\in [0,1]}\Lambda_{R}(b),
   \end{align}
   and
 \begin{align}\label{h10}
   \nonumber\sup_{X\in U_{L}^{\Psi}\cap V^{\Psi}(\mu_{\Psi},\sigma_{\Psi})}\int_{-\infty}^{+\infty}\psi(x)g(\overline{F}_{X}(x))\mathrm{d}x &=\sup_{X\in U_{L}^{\Psi}\cap V^{\Psi}(\mu_{\Psi},\sigma_{\Psi})}\int_{0}^{1}\left(\Psi(F_{X}^{-1}(u))-\mu\right)\mathrm{d}\hat{g}(u)\\
   \nonumber&=\sup_{b\in [0,1]}\left[\sigma_{\Psi}\int_{0}^{b}\frac{\sqrt{3}(2u-2b+b^{2})}{\sqrt{b^{3}(4-3b)}}\mathrm{d}\hat{g}(u)+\sigma_{\Psi}\int_{b}^{1}\sqrt{\frac{3b}{4-3b}}\mathrm{d}\hat{g}(u)\right]\\
  &=\sigma_{\Psi}\sup_{b\in [0,1]}\Lambda_{L}(b).
   \end{align}
   Combining Eqs. (\ref{h9}) and (\ref{h10}) we get
    \begin{align*}
    \sup_{X\in (U_{L}^{\Psi}\cup U_{R}^{\Psi})\cap V^{\Psi} (\mu_{\Psi},\sigma_{\Psi})}\int_{-\infty}^{+\infty}\psi(x)g(\overline{F}_{X}(x))\mathrm{d}x=\sigma_{\Psi}\sup_{b\in[0,1]}\{\Lambda_{R}(b),\Lambda_{L}(b)\}.
    \end{align*}
   Since
$$U_{L}^{\Psi}\cap V^{\Psi} (\mu_{\Psi},\sigma_{\Psi})\subseteq V_{U}^{\Psi} (\mu_{\Psi},\sigma_{\Psi}),~ U_{R}^{\Psi}\cap V^{\Psi}(\mu_{\Psi},\sigma_{\Psi})\subseteq V_{U}^{\Psi} (\mu_{\Psi},\sigma_{\Psi}),$$
  we have
   \begin{align*}
   \sigma_{\Psi}\sup_{b\in[0,1]}\{\Lambda_{R}(b),\Lambda_{L}(b)\}&=\sup_{X\in (U_{L}^{\Psi}\cup U_{R}^{\Psi})\cap V^{\Psi} (\mu_{\Psi},\sigma_{\Psi})}\int_{-\infty}^{+\infty}\psi(x)g(\overline{F}_{X}(x))\mathrm{d}x\\
   &\leq \sup_{X\in V_{U}^{\Psi} (\mu_{\Psi},\sigma_{\Psi})}\int_{-\infty}^{+\infty}\psi(x)g(\overline{F}_{X}(x))\mathrm{d}x.
   \end{align*}
Using Lemmas \ref{le.2} and \ref{le.3},
we get
\begin{align}\label{h11}
    \int_{-\infty}^{+\infty}\psi(x)g(\overline{F}_{X}(x))\mathrm{d}x&=\int_{0}^{1}\Psi(F_{X}^{-1+}(u))\mathrm{d}\hat{g}(u)\leq\int_{0}^{1}\Psi(F_{X}^{-1+}(u))\mathrm{d}\hat{g}_{\ast}(u).
\end{align}
 Since $\hat{g}_{\ast}$ is a convex function in $[0,1]$, there exists a sequence of piecewise linear functions $\tilde{g}_{1}(t) \leq \tilde{g}_{2}(t) \leq \cdots \tilde{g}_{n}(t) \leq \cdots \leq \hat{g}_{\ast}(t)$ such that $\hat{g}_{\ast}(t) = \lim_{n\rightarrow\infty} \tilde{g}_{n}(t)$, where $\tilde{g}_{n}(u)=\sum_{i=1}^{m(n)}(a_{i}^{(n)}u+b_{i}^{(n)})\mathbb{I}_{[t_{i-1}^{(n)},t_{i}^{(n)})}(u)$.
 From the monotone convergence theorem we obtain  $\lim_{n\rightarrow\infty}\int_{0}^{1}\Psi(F_{X}^{-1+}(u))\mathrm{d}\tilde{g}_{n}(u)=\int_{0}^{1}\Psi(F_{X}^{-1+}(u))\mathrm{d}\hat{g}_{\ast}(u)$.
 By the monotonicity of $\tilde{g}_{n}(X)$, combining Eq. (\ref{h11}), we have
\begin{align*}
 & \sup_{X\in V_{U}^{\Psi} (\mu_{\Psi},\sigma_{\Psi})}\int_{-\infty}^{+\infty}\psi(x)g(\overline{F}_{X}(x))\mathrm{d}x\\
  &\leq \sup_{X\in V_{U}^{\Psi} (\mu_{\Psi},\sigma_{\Psi})}\int_{0}^{1}\Psi(F_{X}^{-1+}(u))\mathrm{d}\hat{g}_{\ast}(u)\\
   &=\sup_{X\in V_{U}^{\Psi} (\mu_{\Psi},\sigma_{\Psi})}\lim_{n\rightarrow\infty}\int_{0}^{1}\Psi(F_{X}^{-1+}(u))\mathrm{d}\tilde{g}_{n}(u)\\
   &=\sup_{X\in V_{U}^{\Psi} (\mu_{\Psi},\sigma_{\Psi})}\lim_{n\rightarrow\infty}\int_{0}^{1}\Psi(F_{X}^{-1+}(u))\tilde{g}'_{n}(u)\mathrm{d}u\\
   &=\sup_{X\in V_{U}^{\Psi} (\mu_{\Psi},\sigma_{\Psi})}\lim_{n\rightarrow\infty}\left[\mu_{\Psi} \tilde{g}'_{n}(0)+\int_{0}^{1}\int_{u}^{1}\Psi(F_{X}^{-1+}(t))\mathrm{d}t\mathrm{d}\tilde{g}'_{n}(u)\right]\\
   &\leq\lim_{n\rightarrow\infty}\left[\mu_{\Psi} \tilde{g}'_{n}(0)+\int_{0}^{1}\sup_{X\in V_{U}^{\Psi} (\mu_{\Psi},\sigma_{\Psi})}(1-u)\mathrm{TVaR}_{u}(\Psi(X))\mathrm{d}\tilde{g}'_{n}(u)\right]\\
   &=\mu_{\Psi} \hat{g}_{\ast}'(0)+\int_{0}^{1}\sup_{X\in V_{U}^{\Psi} (\mu_{\Psi},\sigma_{\Psi})}(1-u)\mathrm{TVaR}_{u}(\Psi(X))\mathrm{d}\hat{g}_{\ast}'(u).
  \end{align*}
 By using Eq. (\ref{yy4}) we can get the required results.  $\hfill\square$

Let the lifetime of a component be represented by a non-negative random variable $X$, and let the residual life of the component which has survived up to time $t, t > 0,$ be given by $X_{t}=[X-t|X>t]$.  $X_{t}$ plays an important role in  clinical trials, reliability and survival analysis (see Asadi and Zohrevand (2007),  Finkelstein and Vaupel (2015), Misra and Naqvi (2018)). The survival function of $X_{t}$ is
 \begin{align*}
 \overline{F}_{X_{t}}(x)=
 \begin{cases}
         \frac{\overline{F}_{X}(x)}{\overline{F}_{X}(t)},~\mathrm{when} ~x>t,\\
      1              ,~\mathrm{otherwise}.
 \end{cases}
 \end{align*}
 Hence, for any $v \in (0, 1)$, we get $F_{X_{t}}^{-1}(v)=F_{X}^{-1}\left(F_{X}(t)+(1-F_{X}(t))v\right).$

 \begin{corollary}\label{co.4}
Under the conditions in
Theorem \ref{th.4}, we have
 \begin{align*}
   \sigma_{\Psi}\sup_{b\in[0,1]}\{\Lambda_{R}(b),\Lambda_{L}(b)\}&\leq\sup_{X\in V_{U}^{\Psi} (\mu_{\Psi},\sigma_{\Psi})}\int_{-\infty}^{+\infty}\psi(x)g(\overline{F}_{X_{t}}(x))\mathrm{d}x\leq\frac{\sigma_{\Psi}}{3}\Theta,
\end{align*}
where
$$\Lambda_{R}(b)=\frac{2b\hat{g}_{t}(b)+2\int_{b}^{1}u \hat{g}_{t}'(u)\mathrm{d}u}{\sqrt{(1-b)^{3}(1/3+b)}},
~~\Lambda_{L}(b)=\frac{2\sqrt{3}\left(-b\hat{g}_{t}(b)+\int_{0}^{b}u \hat{g}_{t}'(u)\mathrm{d}u\right)}{\sqrt{b^{3}(4-3b)}},$$
$$\Theta=\int_{0}^{1/2}\sqrt{u(8-9u)}\mathrm{d}\hat{g}_{t,\ast}'(u)+\int_{1/2}^{1}\sqrt{(9u-1)(1-u)}\mathrm{d}\hat{g}_{t,\ast}'(u),$$
  $\hat{g}_{t}(u)=-g\left(\frac{1-u}{1-F_{X}(t)}\right)\mathbb{I}_{[F_{X}(t),1]}(u))$, $u\in[0,1]$, and $\hat{g}_{t,\ast}$ is a convex envelope  of $\hat{g}_{t}$.
\end{corollary}

 A random variable $X_{(t)}=[X|X\leq t]$ describes the past lifetime of the system at age $t$ (see Di Crescenzo and Longobardi (2002), Cal\`{\i} et al. (2017)). The distribution function of $X_{(t)}$ is written as
 \begin{align*}
 F_{X_{(t)}}(x)=
 \begin{cases}
         \frac{F_{X}(x)}{F_{X}(t)},~\mathrm{when} ~x\leq t,\\
      1              ,~\mathrm{otherwise}.
 \end{cases}
 \end{align*}
Thus, for any $s \in (0, 1)$, we have $F_{X_{(t)}}^{-1}(s)=F_{X}^{-1}\left(F_{X}(t)s\right).$
\begin{corollary}\label{co.5}
Under the conditions  in
Theorem \ref{th.4},  we have
 \begin{align*}
 \sigma_{\Psi}\sup_{b\in[0,1]}\{\Lambda_{R}(b),\Lambda_{L}(b)\}&\leq\sup_{X\in V_{U}^{\Psi} (\mu_{\Psi},\sigma_{\Psi})}\int_{-\infty}^{+\infty}\psi(x)g(\overline{F}_{X_{(t)}}(x))\mathrm{d}x\leq\frac{\sigma_{\Psi}}{3}\Theta,
\end{align*}
where
$$\Lambda_{R}(b)=\frac{2b\hat{g}_{(t)}(b)+2\int_{b}^{1}u \hat{g}_{(t)}'(u)\mathrm{d}u}{\sqrt{(1-b)^{3}(1/3+b)}},
~~\Lambda_{L}(b)=\frac{2\sqrt{3}\left(-b\hat{g}_{(t)}(b)+\int_{0}^{b}u \hat{g}_{(t)}'(u)\mathrm{d}u\right)}{\sqrt{b^{3}(4-3b)}},$$
$$\Theta=\int_{0}^{1/2}\sqrt{u(8-9u)}\mathrm{d}\hat{g}_{(t),\ast}'(u)+\int_{1/2}^{1}\sqrt{(9u-1)(1-u)}\mathrm{d}\hat{g}_{(t),\ast}'(u),$$
 $\hat{g}_{(t)}(u)=-g\left(\frac{F_{X}(t)-u}{F_{X}(t)}\right)\mathbb{I}_{[0,F_{X}(t)]}(u)$, $u\in[0,1]$, and $\hat{g}_{(t),\ast}$ is a convex envelope  of $\hat{g}_{(t)}$.
\end{corollary}
\begin{theorem}\label{th.5}Let $g\in\mathcal{G}$, $g$ be continuous with $g(1)=0$, and $\Psi(X)$ be an increasing function with $\psi=\Psi'$.
Then,
 \begin{align}\label{a14}
   \sup_{X\in V_{S}^{\Psi} (\mu_{\Psi},\sigma_{\Psi})}\int_{-\infty}^{+\infty}\psi(x)g(\overline{F}_{X}(x))\mathrm{d}x= \frac{\sigma_{\Psi}}{2}\sqrt{\int_{0}^{1}[(\hat{g}(u)+\hat{g}(1-u))_{\ast}']^{2}\mathrm{d}u},
\end{align}
where $\hat{g}(u)=-g(1-u),~ u \in [0, 1]$.
If $(\hat{g}(u)+\hat{g}(1-u))_{\ast}'=0$ is almost everywhere (a.e.), then (\ref{a14}) can be obtained by any random variable $X \in
V_{S}^{\Psi} (\mu_{\Psi},\sigma_{\Psi})$; If $(\hat{g}(u)+\hat{g}(1-u))_{\ast}'\neq 0 ~(a.e.)$, then (\ref{a14})
 can be obtained by the
worst-case distribution of r.v. $X_{\ast}$ characterized by
$$\Psi(F_{X_{\ast}}^{-1}(u))=\mu_{\Psi}+\sigma_{\Psi}\frac{(\hat{g}(u)+\hat{g}(1-u))_{\ast}'}{\sqrt{\int_{0}^{1}\left((\hat{g}(v)+\hat{g}(1-v))_{\ast}'\right)^{2}\mathrm{d}v}}.$$
\end{theorem}
\noindent $\mathbf{Proof.}$ For $X\in V_{S}^{\Psi} (\mu_{\Psi},\sigma_{\Psi})$, it follows that $\Psi(F_{X}^{-1}(u))+\Psi(F_{X}^{-1}(1-u))=2\mu_{\Psi}~(a.e.)$. Then the weighted entropy can be expressed as
\begin{align*}
\int_{-\infty}^{+\infty}\psi(x)g(\overline{F}_{X}(x))\mathrm{d}x&=\frac{1}{2}\left[\int_{0}^{1}\Psi(F_{X}^{-1}(1-u))\mathrm{d}g(u)+\int_{0}^{1}\Psi(F_{X}^{-1}(u))\mathrm{d}\hat{g}(u)\right]\\
&=\frac{1}{2}\int_{0}^{1}\Psi(F_{X}^{-1}(u))\mathrm{d}(\hat{g}(u)+\hat{g}(1-u)).
\end{align*}
Using same arguments as those in Theorem 3.2 of Zuo and Yin (2025), we can obtain Eq. (\ref{a14}).
$\hfill\square$

\begin{corollary}\label{co.6}
Under the conditions  in
Theorem \ref{th.5}, we have
 \begin{align}\label{z14}
   \sup_{X\in V_{S}^{\Psi} (\mu_{\Psi},\sigma_{\Psi})}\int_{-\infty}^{+\infty}\psi(x)g(\overline{F}_{X_{t}}(x))\mathrm{d}x= \frac{\sigma_{\Psi}}{2}\sqrt{\int_{0}^{1}[(\hat{g}_{t}(u)+\hat{g}_{t}(1-u))_{\ast}']^{2}\mathrm{d}u},
\end{align}
where $\hat{g}_{t}(u)=-g\left(\frac{1-u}{1-F_{X}(t)}\right)\mathbb{I}_{[F_{X}(t),1]}(u))$, $u\in[0,1]$.
If $(\hat{g}_{t}(u)+\hat{g}_{t}(1-u))_{\ast}'=0$ is almost everywhere (a.e.), then (\ref{z14}) can be obtained by any random variable $X \in
V_{S}^{\Psi} (\mu_{\Psi},\sigma_{\Psi})$; If $(\hat{g}_{t}(u)+\hat{g}_{t}(1-u))_{\ast}'\neq0 ~(a.e.)$, then (\ref{z14})
 can be obtained by the
worst-case distribution of r.v. $X_{\ast}$ characterized by
$$\Psi(F_{X_{\ast}}^{-1}(u))=\mu_{\Psi}+\sigma_{\Psi}\frac{(\hat{g}_{t}(u)+\hat{g}_{t}(1-u))_{\ast}'}{\sqrt{\int_{0}^{1}\left((\hat{g}_{t}(v)+\hat{g}_{t}(1-v))_{\ast}'\right)^{2}\mathrm{d}v}}.$$
\end{corollary}
\begin{corollary}\label{co.7}
Under the conditions  in
Theorem \ref{th.5}, we have
 \begin{align}\label{z15}
   \sup_{X\in V_{S}^{\Psi} (\mu_{\Psi},\sigma_{\Psi})}\int_{-\infty}^{+\infty}\psi(x)g(\overline{F}_{X_{(t)}}(x))\mathrm{d}x= \frac{\sigma_{\Psi}}{2}\sqrt{\int_{0}^{1}[(\hat{g}_{(t)}(u)-\hat{g}_{(t)}(1-u))_{\ast}']^{2}\mathrm{d}u},
\end{align}
where
 $\hat{g}_{(t)}(u)=-g\left(\frac{F_{X}(t)-u}{F_{X}(t)}\right)\mathbb{I}_{[0,F_{X}(t)]}(u)$, $u\in[0,1]$.
 If $(\hat{g}_{(t)}(u)-\hat{g}_{(t)}(1-u))_{\ast}'=0$ is almost everywhere (a.e.), then (\ref{z15}) can be obtained by any random variable $X \in
V_{S}^{\Psi} (\mu_{\Psi},\sigma_{\Psi})$; If $(\hat{g}_{(t)}(u)-\hat{g}_{(t)}(1-u))_{\ast}'\neq0 ~(a.e.)$, then (\ref{z15})
 can be obtained by the
worst-case distribution of r.v. $X_{\ast}$ characterized by
$$\Psi(F_{X_{\ast}}^{-1}(u))=\mu_{\Psi}+\sigma_{\Psi}\frac{(\hat{g}_{(t)}(u)-\hat{g}_{(t)}(1-u))_{\ast}'}{\sqrt{\int_{0}^{1}\left((\hat{g}_{(t)}(v)-\hat{g}_{(t)}(1-v))_{\ast}'\right)^{2}\mathrm{d}v}}.$$
\end{corollary}

To derive lower and upper bound of worst-case weighted entropy for symmetric unimodal distributions, we define the set $V(\xi)$ $(\xi\in[0,1])$ of random variables $X$ characterized by (similar to Zhao et al. (2024))
\begin{align*}
\Psi(F_{X}^{-1}(u))=
\begin{cases}
a + c(u - 1 + b),~\mathrm{for}~u\in(0,1-b),\\
a,~\mathrm{for}~u\in[1-b,b],\\
a + c(u - b),~\mathrm{for}~u\in(b,1),
\end{cases}
\end{align*}
where $(a,b,c)\in\mathbb{R} \times[\frac{1}{2}, \max(\xi, 1 - \xi)]\times \mathbb{R}_{+}$.
\begin{theorem}\label{th.6}Let $g\in\mathcal{G}$ with $g(1)=0$, and $\Psi$ be an increasing function such that $\Psi'(x)=\psi(x)$ is the weighted function.  Then,
 \begin{align*}
   \sigma_{\Psi}\sup_{b\in[\frac{1}{2},1]}\Upsilon(b)&\leq\sup_{X\in V_{SU}^{\Psi} (\mu_{\Psi},\sigma_{\Psi})}\int_{-\infty}^{+\infty}\psi(x)g(\overline{F}_{X}(x))\mathrm{d}x\leq \sigma_{\Psi}\Delta,
   \end{align*}
where
$$\Upsilon(b)=\frac{(b-1)\hat{g}(1-b)+b \hat{g}(b)+\int_{0}^{1-b}u \mathrm{d}\hat{g}(u)+\int_{b}^{1}u\mathrm{d}\hat{g}(u)}{\sqrt{2/3(1-b)^{3}}},$$
$$\Delta=\frac{2}{3}\int_{0}^{\frac{1}{3}}\sqrt{u}\mathrm{d}\hat{g}_{\ast}'(u)+\sqrt{3}\int_{\frac{1}{3}}^{\frac{2}{3}}(1-u)u\mathrm{d}\hat{g}_{\ast}'(u)+\frac{2}{3}\int_{\frac{2}{3}}^{1}\sqrt{1-u}\mathrm{d}\hat{g}_{\ast}'(u),$$
 $\hat{g}(u)=-g(1-u),~ u \in [0, 1]$, and $\hat{g}_{\ast}$ be a convex envelope  of $\hat{g}$.
\end{theorem}
\noindent $\mathbf{Proof.}$ By definitions of $V(\xi)$ and $V^{^{\Psi}}(\mu_{\Psi},\sigma_{\Psi})$, we have
\begin{align*}
V(\xi)\cap V^{\Psi}(\mu_{\Psi},\sigma_{\Psi})=\left\{X:~\Psi(F_{X}^{-1}(u))=
\begin{cases}
\mu_{\Psi}+\sigma_{\Psi}\frac{u-1+b}{\sqrt{2/3(1-b)^3}},~\mathrm{for}~u\in(0,1-b),\\
\mu_{\Psi},~\mathrm{for}~u\in[1-b,b],\\
\mu_{\Psi}+\sigma_{\Psi}\frac{u-b}{\sqrt{2/3(1-b)^{3}}},~\mathrm{for}~u\in(b,1),
\end{cases}~where~ b\in[\frac{1}{2}, \max(\xi, 1 - \xi)], ~\xi\in[0,1]\right\}.
\end{align*}
 Thus,
\begin{align*}
   \nonumber\sup_{X\in V(\xi)\cap V^{\Psi}(\mu_{\Psi},\sigma_{\Psi})}\int_{-\infty}^{+\infty}\psi(x)g(\overline{F}_{X}(x))\mathrm{d}x&=\sup_{X\in V(\xi)\cap V^{\Psi}(\mu_{\Psi},\sigma_{\Psi})}\int_{0}^{1}\left(\Psi(F_{X}^{-1}(u))-\mu_{\Psi}\right)\mathrm{d}\hat{g}(u)\\
   \nonumber&=\sup_{b\in[\frac{1}{2},1)}\left\{\sigma_{\Psi}\int_{0}^{1-b}\frac{u-1+b}{\sqrt{2/3(1-b)^{3}}}\mathrm{d}\hat{g}(u)+\sigma_{\Psi}\int_{b}^{1}\frac{u-b}{\sqrt{2/3(1-b)^{3}}}\mathrm{d}\hat{g}(u)\right\}\\
   &=\sigma_{\Psi}\sup_{b\in[\frac{1}{2},1)}\Upsilon(b).
   \end{align*}
   Since
$$V(\xi)\cap V^{\Psi}(\mu_{\Psi},\sigma_{\Psi})\subseteq V_{SU}^{\Psi} (\mu_{\Psi},\sigma_{\Psi}),$$
  we have
   $$\sigma_{\Psi}\sup_{b\in[\frac{1}{2},1)}\Upsilon(b)= \sup_{X\in V(\xi)\cap V^{\Psi} (\mu_{\Psi},\sigma_{\Psi})}\int_{-\infty}^{+\infty}\psi(x)g(\overline{F}_{X}(x))\mathrm{d}x\leq \sup_{X\in V_{SU}^{\Psi} (\mu_{\Psi},\sigma_{\Psi})}\int_{-\infty}^{+\infty}\psi(x)g(\overline{F}_{X}(x))\mathrm{d}x.$$
 Next, using the similar argument as that of Theorem \ref{th.4}
  we can obtain the desired results.
$\hfill\square$

 \begin{corollary}\label{co.8}
Under the conditions in
Theorem \ref{th.6}, we have
 \begin{align*}
   \sigma_{\Psi}\sup_{b\in[\frac{1}{2},1]}\Upsilon(b)&\leq\sup_{X\in V_{SU}^{\Psi} (\mu_{\Psi},\sigma_{\Psi})}\int_{-\infty}^{+\infty}\psi(x)g(\overline{F}_{X_{t}}(x))\mathrm{d}x\leq\sigma_{\Psi}\Delta,
\end{align*}
where
$$\Upsilon(b)=\frac{(b-1)\hat{g}_{t}(1-b)+b \hat{g}_{t}(b)+\int_{0}^{1-b}u \hat{g}_{t}'(u)\mathrm{d}u+\int_{b}^{1}u\hat{g}_{t}'(u)\mathrm{d}u}{\sqrt{2/3(1-b)^{3}}},$$
$$\Delta=\frac{2}{3}\int_{0}^{\frac{1}{3}}\sqrt{u}\mathrm{d}\hat{g}_{t,\ast}'(u)+\sqrt{3}\int_{\frac{1}{3}}^{\frac{2}{3}}(1-u)u\mathrm{d}\hat{g}_{t,\ast}'(u)+\frac{2}{3}\int_{\frac{2}{3}}^{1}\sqrt{1-u}\mathrm{d}\hat{g}_{t,\ast}'(u),$$
 $\hat{g}_{t}(u)=-g\left(\frac{1-u}{1-F_{X}(t)}\right)\mathbb{I}_{[F_{X}(t),1]}(u))$, $u\in[0,1]$, and $\hat{g}_{t,\ast}$ is a convex envelope  of $\hat{g}_{t}$.
\end{corollary}
\begin{corollary}\label{co.9}
Under the conditions in
Theorem \ref{th.6},  we have
 \begin{align*}
   \sigma_{\Psi}\sup_{b\in[\frac{1}{2},1]}\Upsilon(b)&\leq\sup_{X\in V_{SU}^{\Psi} (\mu_{\Psi},\sigma_{\Psi})}\int_{-\infty}^{+\infty}\psi(x)g(\overline{F}_{X_{(t)}}(x))\mathrm{d}x\leq\sigma_{\Psi}\Delta,
   \end{align*}
where
$$\Upsilon(b)=\frac{(b-1)\hat{g}_{(t)}(1-b)+b \hat{g}_{(t)}(b)+\int_{0}^{1-b}u \hat{g}_{(t)}'(u)\mathrm{d}u+\int_{b}^{1}u\hat{g}_{(t)}'(u)\mathrm{d}u}{\sqrt{2/3(1-b)^{3}}},$$
$$\Delta=\frac{2}{3}\int_{0}^{\frac{1}{3}}\sqrt{u}\mathrm{d}\hat{g}_{(t),\ast}'(u)+\sqrt{3}\int_{\frac{1}{3}}^{\frac{2}{3}}(1-u)u\mathrm{d}\hat{g}_{(t),\ast}'(u)+\frac{2}{3}\int_{\frac{2}{3}}^{1}\sqrt{1-u}\mathrm{d}\tilde{g}_{(t),\ast}'(u),$$
 $\hat{g}_{(t)}(u)=-g\left(\frac{F_{X}(t)-u}{F_{X}(t)}\right)\mathbb{I}_{[0,F_{X}(t)]}(u)$, $u\in[0,1]$, and $\hat{g}_{(t),\ast}$ is a convex envelope  of $\hat{g}_{(t)}$.
\end{corollary}
\section{Applications to (weighted) entropies}
In this section, we give applications to some (weighted) entropies, including cumulative Tsallis past entropy, cumulative residual Tsallis entropy of order $\alpha$, extended Gini coefficient, Gini, fractional generalized cumulative residual entropy, cumulative residual entropy,  fractional generalized cumulative entropy, and cumulative entropy. Only the cumulative Tsallis past entropy and fractional generalized
cumulative residual entropy are given here, and the rest is included in the Supplementary Material.
\begin{example}\label{ex.1} Let $g(u)=\frac{1}{\alpha-1}[(1-u)-(1-u)^{\alpha}], ~\alpha>0,~\alpha\neq1,~u\in[0,1]$ in Eq. (\ref{a2}), it is called as the cumulative Tsallis past entropy of a random variable $X$, with distribution function $F_{X}(x)$, denoted by $\mathrm{CT}_{\alpha}(X)$ (when $X\in L^{0}_{+}$, see Cal\`{\i} et al. (2017)), defined by
\begin{align*} \mathrm{CT}_{\alpha}(X)&=\int_{-\infty}^{+\infty}\frac{1}{\alpha-1}[F_{X}(x)-(F_{X}(x))^{\alpha}]\mathrm{d}x.
\end{align*}
In this case, $\hat{g}_{\ast}(u)=\hat{g}(u)=\frac{1}{1-\alpha}(u-u^{\alpha})$, for examples, let $\alpha=\frac{3}{4},2$ and $5$, the graphs of the functions $\hat{g}$ are displayed
in Fig. 1.\\

(i) If $X\in V_{U} (\mu,\sigma)$, by Corollary \ref{co.1}, then
\begin{align*}
  & \sigma\sup_{b\in[0,1]}\{\Lambda_{R}(b),\Lambda_{L}(b)\}\leq\sup_{X\in V_{U} (\mu,\sigma)}\mathrm{CT}_{\alpha}(X)
   \leq\frac{\alpha\sigma}{3}\left[\int_{0}^{1/2}\sqrt{8-9u}u^{\alpha-3/2}\mathrm{d}u+\int_{1/2}^{1}\sqrt{(9u-1)(1-u)}u^{\alpha-2}\mathrm{d}u\right],
\end{align*}
where
$$\Lambda_{R}(b)=\frac{b^{2}+1-2(b^{\alpha+1}+\alpha)/(\alpha+1)}{(1-\alpha)\sqrt{(1-b)^{3}(1/3+b)}},
~~\Lambda_{L}(b)=\frac{\sqrt{3}\left(-b+2b^{\alpha}/(\alpha+1)\right)}{(1-\alpha)\sqrt{b(4-3b)}}.$$
Note that
$$\sup_{b\in[0,1]}\{\Lambda_{R}(b)\}=\max\left\{\frac{\sqrt{3}}{\alpha+1},0,\widetilde{\Lambda}_{R}(b_{0})\right\},$$
and
$$\sup_{b\in[0,1]}\{\Lambda_{L}(b)\}=\max\left\{0,\frac{\sqrt{3}}{\alpha+1},\widetilde{\Lambda}_{L}(b_{1})\right\},~\alpha>\frac{1}{2},$$
where
\begin{align*}
 \widetilde{\Lambda}_{R}(b_{0})=
\begin{cases}
\Lambda_{R}(b_{0}),~\mathrm{if}~\mathrm{exists} ~b_{0},\\
0,~\mathrm{otherwise},
\end{cases}
\widetilde{\Lambda}_{L}(b_{1})=
\begin{cases}
\Lambda_{L}(b_{1}),~\mathrm{if}~\mathrm{exists}~ b_{1},\\
0,~\mathrm{otherwise},
\end{cases}
\end{align*}
$b_{0}\in(0,1)$ and $b_{1}\in(0,1)$ are the solutions to the following equations, respectively (for instance, when $\alpha=2/3,2,5$, $M_{\alpha}(b_{0})$ and $H_{\alpha}(b_{1})$ are shown in Fig. 2 (a-c) and Fig. 3 (a-c), respectively.):
$$M_{\alpha}(b_{0})=3(\alpha-1)b^{\alpha+1}_{0}-2(\alpha+1)b^{\alpha}_{0}-(\alpha+1)b^{\alpha-1}_{0}+2(\alpha+1)b_{0}+4-2\alpha=0,$$
$$H_{\alpha}(b_{1})=3(\alpha-1)b_{1}^{\alpha}+(4\alpha-2)b_{1}^{\alpha-1}-(\alpha+1)=0.$$
Therefore,
\begin{align*}
\sigma\max\left\{\frac{\sqrt{3}}{\alpha+1},\widetilde{\Lambda}_{R}(b_{0}),\widetilde{\Lambda}_{L}(b_{1})\right\}\leq\sup_{X\in V_{U} (\mu,\sigma)}\mathrm{CT}_{\alpha}(X),~\alpha>\frac{1}{2}
.
\end{align*}

\noindent(ii) If $X\in V_{S} (\mu,\sigma)$, by Corollary \ref{co.2}, then
\begin{align}\label{s5}
   \sup_{X\in V_{S} (\mu,\sigma)}\mathrm{CT}_{\alpha}(X)= \frac{\alpha\sigma}{\sqrt{2}|1-\alpha|}\sqrt{\frac{1}{2\alpha-1}-\mathrm{B}(\alpha,\alpha)},~\alpha>\frac{1}{2},
   \end{align}
which can be obtained by the
worst-case distribution of r.v. $X_{\ast}$ characterized by
$$F_{X_{\ast}}^{-1}(u)=\mu+\sigma\frac{\mathrm{sign}(1-\alpha)[(1-u)^{\alpha-1}-u^{\alpha-1}]}{\sqrt{2\left(\frac{1}{2\alpha-1}-\mathrm{B}(\alpha,\alpha)\right)}}.$$
where $|\cdot|$, $\mathrm{B}(\cdot,\cdot)$ and $\mathrm{sign}(\cdot)$ are the absolute value, Beta and sign functions, respectively.\\
(iii) If $X\in V_{SU} (\mu,\sigma)$, by Corollary \ref{co.3}, then
\begin{align*}
  &\sigma\sup_{b\in[\frac{1}{2},1)}\Upsilon(b)\leq \sup_{X\in V_{SU} (\mu,\sigma)}\mathrm{CT}_{\alpha}(X)\\
  &\leq\alpha\sigma\left[\frac{4}{3^{\alpha+1/2}(2\alpha-1)}+\sqrt{3}\left[\mathbf{B}_{\frac{2}{3}}(\alpha-\frac{1}{2},2)-\mathbf{B}_{\frac{1}{3}}(\alpha-\frac{1}{2},2)\right]+\frac{2}{3}\left[\mathbf{B}(\alpha-1,\frac{3}{2})-\mathbf{B}_{\frac{2}{3}}(\alpha-1,\frac{3}{2})\right] \right] ,~\alpha>1,
\end{align*}
where
$$\Upsilon(b)=\frac{b^{\alpha+1}-(1-b)^{\alpha+1}-(\alpha+1)b+\alpha}{(\alpha^{2}-1)\sqrt{2/3(1-b)^{3}}}.$$
Note that
$$\sup_{b\in[\frac{1}{2},1)}\Upsilon(b)=\max\left\{\frac{\sqrt{3}}{\alpha+1},0,\widetilde{\Upsilon}(b_{2})\right\},~\alpha>\frac{1}{2},$$
where
 \begin{align*}
 \widetilde{\Upsilon}(b_{2})=
\begin{cases}
\Upsilon(b_{2}),~\mathrm{if}~\mathrm{exists} ~b_{2},\\
0,~\mathrm{otherwise},
\end{cases}
\end{align*}
$b_{2}\in(\frac{1}{2},1)$ is the solution to the following equation (for example, when $\alpha=2/3,2,5$, $K_{\alpha}(b_{2})$ are presented in Fig. 4 (a-c), respectively.):
$$K_{\alpha}(b_{2})=(1-2\alpha)b_{2}^{\alpha+1}+(2\alpha-1)(1-b_{2})^{\alpha+1}+2(\alpha+1) b_{2}^{\alpha}-(\alpha+1)b_{2}+\alpha-2=0.$$
Thus,
\begin{align*}
   \sigma\max\left\{\frac{\sqrt{3}}{\alpha+1},\widetilde{\Upsilon}(b_{2})\right\}\leq \sup_{X\in V_{SU} (\mu,\sigma)}\mathrm{CT}_{\alpha}(X),~\alpha>\frac{1}{2}.
\end{align*}
\end{example}

Letting $\alpha = 2$ in Example \ref{ex.1}, $\mathrm{CT}_{\alpha}(X)$ will be Gini mean semi-difference ($\mathcal{G}ini(X)$) (see, e.g., Hu and Chen (2020), Yin et al. (2023)), defined by
\begin{align*}
\mathcal{G}ini(X)&=\int_{-\infty}^{+\infty}[F_{X}(x)-(F_{X}(x))^{2}]\mathrm{d}x.
\end{align*}
(i) If $X\in V_{U} (\mu,\sigma)$, then
\begin{align*}
\frac{\sigma}{\sqrt{3}}\leq\sup_{X\in V_{U} (\mu,\sigma)}\mathcal{G}ini(X)\leq\frac{2\sigma}{3}\left[\int_{0}^{1/2}\sqrt{u(8-9u)}\mathrm{d}u+\int_{1/2}^{1}\sqrt{(9u-1)(1-u)}\mathrm{d}u\right].
\end{align*}
(ii) If $X\in V_{S} (\mu,\sigma)$, then
\begin{align*}
   \sup_{X\in V_{S} (\mu,\sigma)}\mathcal{G}ini(X)=\frac{\sigma}{\sqrt{3}},
\end{align*}
which can be obtained by the
worst-case distribution of r.v. $X_{\ast}$ characterized by
$$F_{X_{\ast}}^{-1}(u)=\mu+\sigma\sqrt{3}(2u-1).$$
(iii) If $X\in V_{SU} (\mu,\sigma)$, then
\begin{align*}
   \frac{\sigma}{\sqrt{3}}\leq\sup_{X\in V_{SU} (\mu,\sigma)}\mathcal{G}ini(X)\leq2\sigma\left[\sqrt{3}\left(\mathbf{B}_{\frac{2}{3}}(\frac{3}{2},2)-\mathbf{B}_{\frac{1}{3}}(\frac{3}{2},2)\right)+\frac{2}{3}\left(\mathbf{B}(1,\frac{3}{2})-\mathbf{B}_{\frac{2}{3}}(1,\frac{3}{2})\right)+\frac{4}{27\sqrt{3}}\right].
\end{align*}
\begin{remark}In Zuo and Yin (2025), authors has derived that $\sup_{X\in V (\mu,\sigma)}\mathcal{G}ini(X)=\frac{\sigma}{\sqrt{3}}$. Since $\sup_{X\in V_{SU} (\mu,\sigma)}\mathcal{G}ini(X)\leq\sup_{X\in V_{U} (\mu,\sigma)}\mathcal{G}ini(X)\leq\sup_{X\in V (\mu,\sigma)}\mathcal{G}ini(X)$, combining the above result $\frac{\sigma}{\sqrt{3}}\leq\sup_{X\in V_{SU} (\mu,\sigma)}\mathcal{G}ini(X)$, we can conclude that
$$\sup_{X\in V_{SU} (\mu,\sigma)}\mathcal{G}ini(X)=\sup_{X\in V_{U} (\mu,\sigma)}\mathcal{G}ini(X)=\frac{\sigma}{\sqrt{3}}.$$
\end{remark}
\begin{example}\label{ex.4} Let $g(u)=\frac{1}{\Gamma(\alpha+1)}u[-\log(u)]^{\alpha}, ~\alpha>0,~u\in[0,1]$ in Eq. (\ref{a2})
, it is called as the  fractional generalized cumulative residual entropy of
a random variable $X$, with distribution function $F_{X}(x)$ and tail distribution function $\overline{F}_{X}(x)$, denoted by $FGR\mathcal{E}_{\alpha}(X)$ (when $X\in(0,c)$, see Di Crescenzo et al. (2021)), defined as
\begin{align*}
 FGR\mathcal{E}_{\alpha}(X)& =\frac{1}{\Gamma(\alpha+1)}\int_{-\infty}^{+\infty}\overline{F}_{X}(x)\left[-\log\left(\overline{F}_{X}(x)\right)\right]^{\alpha}\mathrm{d}x.
 \end{align*}
 In this case, $\hat{g}(u)=-\frac{1}{\Gamma(\alpha+1)}(1-u)[-\log(1-u)]^{\alpha}$, when $\alpha\in(0,1]$, $\hat{g}_{\ast}(u)=\hat{g}(u)$; when $\alpha\in(1,+\infty]$,
\begin{align*}
 \hat{g}_{\ast}(u)=[-g(1-u)]_{\ast}=
 \begin{cases}
 b u,~u\in[0,u_{0}),\\
 \frac{-1}{\Gamma(\alpha+1)}(1-u)[-\log(1-u)]^{\alpha},~u\in[u_{0},1],
 \end{cases}
 \end{align*}
 where $b=\frac{-1}{u_{0}\Gamma(\alpha+1)}(1-u_{0})[-\log(1-u_{0})]^{\alpha}$, and $u_{0}\in[1-e^{1-\alpha},1]$ is the solution to the following equation: $\alpha u_{0}+\log(1-u_{0})=0$. For examples, let $\alpha=3/4,~1~\mathrm{and} ~2$, the graphs of the function $\hat{g}$ are shown
in Fig. 5 (a). Let $\alpha=2$, the graphs of the functions $\hat{g}$ and $\hat{g}_{\ast}$ are displayed
in Fig. 5 (b).\\
(i) If $X\in V_{U} (\mu,\sigma)$, by Corollary \ref{co.1}
, then\\
(i-1) When $\alpha\in(0,1)$,
\begin{align*}
 &\sigma\sup_{b\in[0,1]}\{\Lambda_{R}(b),\Lambda_{L}(b)\}\leq  \sup_{X\in V_{U} (\mu,\sigma)}FGR\mathcal{E}_{\alpha}(X)
 \leq\frac{\alpha\sigma}{3\Gamma(\alpha+1)}\left[\int_{0}^{1/2}\sqrt{u(8-9u)}w_{1}(u)\mathrm{d}u+\int_{1/2}^{1}\sqrt{9u-1}w_{2}(u)\mathrm{d}u\right],
\end{align*}
(i-2) When $\alpha\in(1,+\infty)$, if $u_{0}\in(0,\frac{1}{2}),$
\begin{align*}
 &\sigma\sup_{b\in[0,1]}\{\Lambda_{R}(b),\Lambda_{L}(b)\}\leq  \sup_{X\in V_{U} (\mu,\sigma)}FGR\mathcal{E}_{\alpha}(X)
 \leq\frac{\alpha\sigma}{3\Gamma(\alpha+1)}\left[\int_{u_{0}}^{1/2}\sqrt{u(8-9u)}w_{1}(u)\mathrm{d}u+\int_{1/2}^{1}\sqrt{9u-1}w_{2}(u)\mathrm{d}u\right],
\end{align*}
if $u_{0}\in[\frac{1}{2},1),$
\begin{align*}
 &\sigma\sup_{b\in[0,1]}\{\Lambda_{R}(b),\Lambda_{L}(b)\}\leq  \sup_{X\in V_{U} (\mu,\sigma)}FGR\mathcal{E}_{\alpha}(X)
 \leq\frac{\alpha\sigma}{3\Gamma(\alpha+1)}\left[\int_{u_{0}}^{1}\sqrt{9u-1}w_{2}(u)\mathrm{d}u\right],
\end{align*}
where
$w_{1}(u)=\frac{[-\log(1-u)]^{\alpha-1}-(\alpha-1)[-\log(1-u)]^{\alpha-2}}{1-u}$, $w_{2}(u)=\frac{[-\log(1-u)]^{\alpha-1}-(\alpha-1)[-\log(1-u)]^{\alpha-2}}{\sqrt{1-u}}$,
$$\Lambda_{R}(b)=\frac{[-\log(1-b)]^{\alpha}(1-b)^{2}+\frac{\alpha}{2^{\alpha}}\Gamma(\alpha,-2\log(1-b))}{\Gamma(\alpha+1)\sqrt{(1-b)^{3}(1/3+b)}},$$
$$\Lambda_{L}(b)=\frac{\sqrt{3}\left\{-[-\log(1-b)]^{\alpha}(1-b)^{2}+\frac{\alpha}{2^{\alpha}}\boldsymbol{\gamma}(\alpha,-2\log(1-b))\right\}}{\Gamma(\alpha+1)\sqrt{b^{3}(4-3b)}},$$
and $\Gamma(\cdot,\cdot)$ and $\boldsymbol{\gamma}(\cdot,\cdot)$ denote upper and lower incomplete gamma functions, respectively (see, e.g., Abramowitz and Stegun (1965), Chapter 6), i.e.,
$$\Gamma(s,x)=\int_{x}^{+\infty}t^{s-1}\mathrm{e}^{-t}\mathrm{d}t,~~\boldsymbol{\gamma}(s,x)=\int_{0}^{x}t^{s-1}\mathrm{e}^{-t}\mathrm{d}t.$$
Note that
$$\sup_{b\in[0,1]}\{\Lambda_{R}(b)\}=\max\left\{\frac{\sqrt{3}}{2^{\alpha}},0,\widetilde{\Lambda}_{R}(b_{0})\right\},~
\mathrm{and}
~\sup_{b\in[0,1]}\{\Lambda_{L}(b)\}=\max\left\{0,\frac{\sqrt{3}}{2^{\alpha}},\widetilde{\Lambda}_{L}(b_{1})\right\},$$
where
\begin{align*}
 \widetilde{\Lambda}_{R}(b_{0})=
\begin{cases}
\Lambda_{R}(b_{0}),~\mathrm{if}~\mathrm{exists} ~b_{0},\\
0,~\mathrm{otherwise},
\end{cases}
\widetilde{\Lambda}_{L}(b_{1})=
\begin{cases}
\Lambda_{L}(b_{1}),~\mathrm{if}~\mathrm{exists}~ b_{1},\\
0,~\mathrm{otherwise},
\end{cases}
\end{align*}
$b_{0}\in(0,1)$ and $b_{1}\in(0,1)$ are the solutions to the following equations, respectively:
$$\frac{\alpha}{2^{\alpha}}b_{0}\Gamma(\alpha,-2\log(1-b_{0}))-\frac{1}{3}[-\log(1-b_{0})]^{\alpha}(1-b_{0})^{2}=0,$$
$$(3-2b_{1})[-\log(1-b_{1})]^{\alpha}-\frac{3\alpha}{2^{\alpha}}\boldsymbol{\gamma}(\alpha,-2\log(1-b_{1}))=0.$$
Therefore,
\begin{align*}
   \sup_{X\in V_{U} (\mu,\sigma)}FGR\mathcal{E}_{\alpha}(X)\geq \sigma\max\left\{\frac{\sqrt{3}}{2^{\alpha}},\widetilde{\Lambda}_{R}(b_{0}),\widetilde{\Lambda}_{L}(b_{1})\right\}.
\end{align*}
(ii) If $X\in V_{S} (\mu,\sigma)$, by Corollary \ref{co.2}
, then\\
(ii-1) When $\alpha\in(0,1]$ (For $X \in (0, c)$ and $\alpha\in(0,1]$, see Xiong et al. (2019)),
 \begin{align*}
   \sup_{X\in V_{S} (\mu,\sigma)}FGR\mathcal{E}_{\alpha}(X)= \frac{\sigma}{\sqrt{2}\Gamma(\alpha+1)}\sqrt{\alpha^{2}\Gamma(2\alpha-1)-\eta},
\end{align*}
which can be obtained by the
worst-case distribution of r.v. $X_{\ast}$ characterized by
$$F_{X_{\ast}}^{-1}(u)=\mu+\sigma\frac{[-\log(u)]^{\alpha-1}[\log(u)+\alpha]-[-\log(1-u)]^{\alpha-1}[\log(1-u)+\alpha]}{\sqrt{2}\sqrt{\alpha^{2}\Gamma(2\alpha-1)-\eta}},$$
where
$$\eta=\int_{0}^{1}[\log(u)]^{\alpha}[\log(1-u)]^{\alpha}\mathrm{d}u+2\alpha\int_{0}^{1}[\log(u)]^{\alpha}[\log(1-u)]^{\alpha-1}\mathrm{d}u+\alpha^{2}\int_{0}^{1}[\log(u)]^{\alpha-1}[\log(1-u)]^{\alpha-1}\mathrm{d}u.$$
(ii-2) When $\alpha\in(1,\infty),$
 further, when $u_{0}\geq\frac{1}{2}$,
\begin{align*}
   &\sup_{X\in V_{S} (\mu,\sigma)}FGR\mathcal{E}_{\alpha}(X)=\\ &\sigma\sqrt{\frac{b^{2}(1-u_{0})}{2}+\frac{[(-\log(1-u_{0}))^{2\alpha}(1-u_{0})+\alpha^{2}\Gamma(2\alpha-1,-\log(1-u_{0}))]}{2\Gamma^{2}(\alpha+1)}-\frac{b[-\log(1-u_{0})]^{\alpha}(1-u_{0})}{\Gamma(\alpha+1)}},
\end{align*}
which can be obtained by the
worst-case distribution of r.v. $X_{\ast}$ characterized by
\begin{align*}
F_{X_{\ast}}^{-1}(u)=
\begin{cases}
\mu+\sigma\frac{b+\frac{1}{\Gamma(\alpha+1)}[-\log(u)]^{\alpha-1}[\log(u)+\alpha]}{\sqrt{2b^{2}(1-u_{0})+\frac{2[(-\log(1-u_{0}))^{2\alpha}(1-u_{0})+\alpha^{2}\Gamma(2\alpha-1,-\log(1-u_{0}))]}{\Gamma^{2}(\alpha+1)}-\frac{4b[-\log(1-u_{0})]^{\alpha}(1-u_{0})}{\Gamma(\alpha+1)}}},~&u\in[0,1-u_{0}),\\
\mu,~&u\in[1-u_{0},u_{0}),\\
\mu-\sigma\frac{b+\frac{1}{\Gamma(\alpha+1)}[-\log(1-u)]^{\alpha-1}[\log(1-u)+\alpha]}{\sqrt{2b^{2}(1-u_{0})+\frac{2[(-\log(1-u_{0}))^{2\alpha}(1-u_{0})+\alpha^{2}\Gamma(2\alpha-1,-\log(1-u_{0}))]}{\Gamma^{2}(\alpha+1)}-\frac{4b[-\log(1-u_{0})]^{\alpha}(1-u_{0})}{\Gamma(\alpha+1)}}},~&u\in[u_{0},1],
\end{cases}
\end{align*}
when $u_{0}<\frac{1}{2}$,
\begin{align}\label{s6}
   \nonumber&\sup_{X\in V_{S} (\mu,\sigma)}FGR\mathcal{E}_{\alpha}(X)=\\ &\sigma\sqrt{\frac{b^{2}u_{0}}{2}-\frac{b[-\log(u_{0})]^{\alpha}u_{0}}{\Gamma(\alpha+1)}+\frac{\alpha^{2}\Gamma(2\alpha-1,-\log(1-u_{0}))+[-\log(1-u_{0})]^{2\alpha}(1-u_{0})}{2\Gamma^{2}(\alpha+1)}-\frac{\delta}{2\Gamma^{2}(\alpha+1)}},
\end{align}
where $$\delta=\int_{u_{0}}^{1-u_{0}}[(\log(u))^{\alpha}(\log(1-u))^{\alpha}+\alpha(\log(u))^{\alpha}(\log(1-u))^{\alpha-1}+\alpha(\log(u))^{\alpha-1}(\log(1-u))^{\alpha}+\alpha^{2}(\log(u))^{\alpha-1}(\log(1-u))^{\alpha-1}]\mathrm{d}u.$$
Eq.(\ref{s6}) can be obtained by the
worst-case distribution of r.v. $X_{\ast}$ characterized by
\begin{align*}
F_{X_{\ast}}^{-1}(u)=
\begin{cases}
\mu+\sigma\frac{b+\frac{1}{\Gamma(\alpha+1)}[-\log(u)]^{\alpha-1}[\log(u)+\alpha]}{\sqrt{2b^{2}u_{0}-\frac{4b[-\log(u_{0})]^{\alpha}u_{0}}{\Gamma(\alpha+1)}+\frac{2[\alpha^{2}\Gamma(2\alpha-1,-\log(1-u_{0}))+[-\log(1-u_{0})]^{2\alpha}(1-u_{0})]}{\Gamma^{2}(\alpha+1)}-\frac{2\delta}{\Gamma^{2}(\alpha+1)}}},~&u\in[0,u_{0}),\\
\mu+\sigma\frac{[-\log(u)]^{\alpha-1}[\log(u)+\alpha]-[-\log(1-u)]^{\alpha-1}[\log(1-u)+\alpha]}{\Gamma(\alpha+1)\sqrt{2b^{2}u_{0}-\frac{4b[-\log(u_{0})]^{\alpha}u_{0}}{\Gamma(\alpha+1)}+\frac{2[\alpha^{2}\Gamma(2\alpha-1,-\log(1-u_{0}))+[-\log(1-u_{0})]^{2\alpha}(1-u_{0})]}{\Gamma^{2}(\alpha+1)}-\frac{2\delta}{\Gamma^{2}(\alpha+1)}}},~&u\in[u_{0},1-u_{0}),\\
\mu-\sigma\frac{b+\frac{1}{\Gamma(\alpha+1)}[-\log(1-u)]^{\alpha-1}[\log(1-u)+\alpha]}{\sqrt{2b^{2}u_{0}-\frac{4b[-\log(u_{0})]^{\alpha}u_{0}}{\Gamma(\alpha+1)}+\frac{2[\alpha^{2}\Gamma(2\alpha-1,-\log(1-u_{0}))+[-\log(1-u_{0})]^{2\alpha}(1-u_{0})]}{\Gamma^{2}(\alpha+1)}-\frac{2\delta}{\Gamma^{2}(\alpha+1)}}},~&u\in[1-u_{0},1].
\end{cases}
\end{align*}
(iii) If $X\in V_{SU} (\mu,\sigma)$, by Corollary \ref{co.3}
, then\\
(iii-1) When $\alpha\in(0,1)$,
\begin{align*}
   & \sigma\sup_{b\in[\frac{1}{2},1)}\Upsilon(b)\leq\sup_{X\in V_{SU} (\mu,\sigma)}FGR\mathcal{E}_{\alpha}(X)
   \leq\frac{\alpha\sigma}{\Gamma(\alpha+1)}\left[\frac{2}{3}\int_{0}^{\frac{1}{3}}\sqrt{u}w_{1}(u)\mathrm{d}u+\sqrt{3}\int_{\frac{1}{3}}^{\frac{2}{3}}uw_{3}(u)\mathrm{d}u+\frac{2}{3}\int_{\frac{2}{3}}^{1}w_{2}(u)\mathrm{d}u\right],
\end{align*}
(iii-2) when $\alpha\in(1,+\infty)$, if $u_{0}\in(0,\frac{1}{3})$,
\begin{align*}
   & \sigma\sup_{b\in[\frac{1}{2},1)}\Upsilon(b)\leq\sup_{X\in V_{SU} (\mu,\sigma)}FGR\mathcal{E}_{\alpha}(X)
   \leq\frac{\alpha\sigma}{\Gamma(\alpha+1)}\left[\frac{2}{3}\int_{u_{0}}^{\frac{1}{3}}\sqrt{u}w_{1}(u)\mathrm{d}u+\sqrt{3}\int_{\frac{1}{3}}^{\frac{2}{3}}uw_{3}(u)\mathrm{d}u+\frac{2}{3}\int_{\frac{2}{3}}^{1}w_{2}(u)\mathrm{d}u\right],
\end{align*}
if $u_{0}\in[\frac{1}{3},\frac{2}{3})$,
\begin{align*}
   & \sigma\sup_{b\in[\frac{1}{2},1)}\Upsilon(b)\leq\sup_{X\in V_{SU} (\mu,\sigma)}FGR\mathcal{E}_{\alpha}(X)
   \leq\frac{\alpha\sigma}{\Gamma(\alpha+1)}\left[\sqrt{3}\int_{u_{0}}^{\frac{2}{3}}uw_{3}(u)\mathrm{d}u+\frac{2}{3}\int_{\frac{2}{3}}^{1}w_{2}(u)\mathrm{d}u\right],
\end{align*}
if $u_{0}\in[\frac{2}{3},1)$,
\begin{align*}
   & \sigma\sup_{b\in[\frac{1}{2},1)}\Upsilon(b)\leq\sup_{X\in V_{SU} (\mu,\sigma)}FGR\mathcal{E}_{\alpha}(X)
   \leq\frac{\alpha\sigma}{\Gamma(\alpha+1)}\left[\frac{2}{3}\int_{u_{0}}^{1}w_{2}(u)\mathrm{d}u\right],
\end{align*}
where $w_{3}(u)=[-\log(1-u)]^{\alpha-1}-(\alpha-1)[-\log(1-u)]^{\alpha-2}$,
$$\Upsilon(b)=\frac{\Gamma(\alpha+1,-2\log(1-b))+\boldsymbol{\gamma}(\alpha+1,-2\log(b))}{2^{\alpha+1}\Gamma(\alpha+1)\sqrt{2/3(1-b)^{3}}}.$$
Note that
$$\sup_{b\in[\frac{1}{2},1)}\Upsilon(b)=\max\left\{\frac{\sqrt{3}}{2^{\alpha}},0,\widetilde{\Upsilon}(b_{2})\right\},$$
where
 \begin{align*}
 \widetilde{\Upsilon}(b_{2})=
\begin{cases}
\Upsilon(b_{2}),~\mathrm{if}~\mathrm{exists} ~b_{2},\\
0,~\mathrm{otherwise},
\end{cases}
\end{align*}
$b_{2}\in(\frac{1}{2},1)$ is the solution to the following equation:
$$2^{\alpha+2}[-\log(1-b_{2})]^{\alpha}(1-b_{2})^{2}+2^{\alpha+2}[-\log(b_{2})]^{\alpha}(1-b_{2})b_{2}-3[\Gamma(\alpha+1,-2\log(1-b_{2}))+\boldsymbol{\gamma}(\alpha+1,-2\log(b_{2}))]=0.$$
Thus,
\begin{align*}
   \sup_{X\in V_{SU} (\mu,\sigma)}FGR\mathcal{E}_{\alpha}(X)\geq \sigma\max\left\{\frac{\sqrt{3}}{2^{\alpha}},\widetilde{\Upsilon}(b_{2})\right\}.
\end{align*}
\end{example}

\begin{remark}
Note that when $\alpha = n \in \mathbb{N}$, $FGR\mathcal{E}_{\alpha}(X)$ will be generalized cumulative residual entropy (see, e.g., Psarrakos and Navarro (2013)).
In particular, when $\alpha= 1$ in Example \ref{ex.4}, $FGR\mathcal{E}_{\alpha}(X)$ will be reduced to cumulative residual entropy ($\mathcal{E}(X)$), defined by (see Rao et al. (2004))
\begin{align*}
\mathcal{E}(X)
&=-\int_{0}^{+\infty}\overline{F}_{X}(x)\mathrm{log}(\overline{F}_{X}(x))\mathrm{d}x.
\end{align*}
In this case, $\hat{g}_{\ast}(u)=\hat{g}(u)=(1-u)\log(1-u)$.\\
(i) If $X\in V_{U} (\mu,\sigma)$, then we have
\begin{align*}
  &0.927341\sigma\approx\frac{[1/2-\log(1-b_{0})]\sqrt{1-b_{0}}}{\sqrt{1/3+b_{0}}}\sigma\leq \sup_{X\in V_{U} (\mu,\sigma)}\mathcal{E}(X)
  \leq \frac{\sigma}{3}\left[\int_{0}^{\frac{1}{2}}\frac{\sqrt{u(8-9u)}}{1-u}\mathrm{d}u+\int_{\frac{1}{2}}^{1}\sqrt{\frac{9u-1}{1-u}}\mathrm{d}u\right],
\end{align*}
where $b_{0}\in(0,1)$ is the solution to the following equation: $\frac{2}{3}\log(1-b_{0})+b_{0}=0$ ($b_{0}\approx0.582812$).\\
(ii) If $X\in V_{S} (\mu,\sigma)$, then we get\\
\begin{align*}
   \sup_{X\in V_{S} (\mu,\sigma)}\mathcal{E}(X)= \frac{\pi\sigma}{2\sqrt{3}},
\end{align*}
which can be obtained by the
worst-case distribution of r.v. $X_{\ast}$ characterized by
$$F_{X_{\ast}}^{-1}(u)=\mu+\sigma\frac{\sqrt{3}[\log(u)-\log(1-u)]}{\pi}.$$
(iii) If $X\in V_{SU} (\mu,\sigma)$,  then
\begin{align*}
 &0.878282\sigma \approx \frac{\Gamma(2,-2\log(1-b_{2}))+\boldsymbol{\gamma}(2,-2\log(b_{2}))}{4\sqrt{2/3(1-b_{2})^{3}}} \sigma\leq \sup_{X\in V_{SU} (\mu,\sigma)}\mathcal{E}(X)
 \leq \sigma\left[\frac{2}{3}\log\left(\frac{\sqrt{3}+1}{\sqrt{3}-1}\right)+\frac{1}{2\sqrt{3}}\right],
\end{align*}
$b_{2}\in(\frac{1}{2},1)$ is the solution to the following equation:
$$8[-\log(1-b_{2})](1-b_{2})^{2}+8[-\log(b_{2})](1-b_{2})b_{2}-3[\Gamma(2,-2\log(1-b_{2}))+\boldsymbol{\gamma}(2,-2\log(b_{2}))]=0 ~(b_{2}\approx0.617477).$$
\end{remark}
\begin{remark}Note that for $X\in L_{+}^{0}$, the corresponding results of weighted cumulative Tsallis entropy of order $\alpha$ (see, e.g., Chakraborty and Pradhan (2023)) can be obtained by replacing $(\mu,\sigma,F^{-1}_{X}(u))$  with $(\mu_{\Psi}=\frac{\mathrm{E}[X^{2})}{2},\sigma_{\Psi}=\frac{\sqrt{\mathrm{Var}(X^{2})}}{2},\Psi(F^{-1}_{X}(u))=\frac{(F^{-1}_{X}(u)))^{2}}{2}$ in Example \ref{ex.1}. The corresponding results of weighted cumulative residual entropy with weight function $\psi$ ($\mathrm{WGCRE}_{\psi}(X)$, see, e.g., Suhov and Yasaei Sekeh (2015)) can be obtained by replacing $(\mu,\sigma,F^{-1}_{X}(u))$  with $(\mu_{\Psi},\sigma_{\Psi},\Psi(F^{-1}_{X}(u))$, and letting $\alpha=1$ in Example \ref{ex.4}. Further, when $\psi(x)= x$, $\mathrm{WGCRE}_{\psi}(X)$ will reduce to weighted cumulative residual entropy ($\mathcal{E}^{w}(X)$), see, e.g., Misagh et al. (2011) and  Mirali et al. (2017).
\end{remark}
\section{Applications to shortfalls}
In this section, we present applications to several shortfalls, such as extended Gini shortfall, Gini shortfall, shortfall of cumulative residual entropy, and shortfall of cumulative residual Tsallis entropy. Only the extended Gini shortfall and Gini shortfall are given here, and the rest is included in the Supplementary Material.
\begin{example}\label{ex.10} Let $\hat{g}(u)=\frac{1}{(1-p)^{2}}\left[(1-p)u+2\tau\left((1-u)^{r}+(1-p)^{r-1}u\right)+p(p-1)-2\tau(1-p)^{r-1}\right]\mathbb{I}_{[p,1]}(u)$, $u\in[0,1]$, $r>1$, $\tau\in[0,1/(2(r-1)(1-p)^{r-2})]$, and $g(1)=\hat{g}(1)=1$ in Lemma \ref{le.1}, it is called as
the extended Gini shortfall (EGS) of a random variable $X$ with parameter $r>1$,
$p\in(0,1)$ and the loading
parameter $\tau \geq 0$,
  denoted by $\mathrm{EGS}_{r,p}^{\tau} (X)$ (see, e.g., Berkhouch et al., 2018), defined as
\begin{align*}
\mathrm{EGS}_{r,p}^{\tau} (X) = \mathrm{ES}_{p}(X) + \tau \mathrm{TEGini}_{r,p}(X),
\end{align*}
 where
 $
 \mathrm{TEGini}_{r,p}(X)=-\int_{x_{p}}^{+\infty}\frac{2}{(1-p)^{2}}\left[\left(\overline{F}_{X}(x)\right)^{r}-(1-p)^{r-1}\overline{F}_{X}(x)\right]\mathrm{d}x
 $
  denotes tail-based extended Gini coefficient with $x_{p}= F^{-1}_{X}(p)$.\\
 (i) If $X\in V_{U} (\mu,\sigma)$, by Theorem \ref{th.1}, then
\begin{align*}
   &\sup_{X\in V_{U} (\mu,\sigma)}\mathrm{EGS}_{r,p}^{\tau} (X) \\
   &= \mu +\frac{2r(r-1)\tau\sigma}{3(1-p)^{2}}\left[\int_{0}^{1/2}\sqrt{u(8-9u)}(1-u)^{r-2}\mathbb{I}_{[p,1]}(u)\mathrm{d}u+\int_{1/2}^{1}\sqrt{9u-1}(1-u)^{r-3/2}\mathbb{I}_{[p,1]}(u)\mathrm{d}u\right],~\tau\neq0.
\end{align*}
(ii) If $X\in V_{S} (\mu,\sigma)$, by Theorem \ref{th.2}, then\\
when $p\geq\frac{1}{2}$,
\begin{align*}
   \sup_{X\in V_{S} (\mu,\sigma)}\mathrm{EGS}_{r,p}^{\tau} (X)= \mu +{\sigma}\sqrt{\frac{1}{2}\left[\frac{1}{1-p}+\frac{4\tau^{2}(r-1)^{2}(1-p)^{2r-5}}{2r-1}\right]},
\end{align*}
which can be obtained by the
worst-case distribution of r.v. $X_{\ast}$ characterized by
\begin{align*}
F_{X_{\ast}}^{-1}(u)=
\begin{cases}
\mu-\sigma\frac{1-p+2\tau[-r u^{r-1}+(1-p)^{r-1}]}{\sqrt{2\left[(1-p)^{3}+\frac{4\tau^{2}(r-1)^{2}(1-p)^{2r-1}}{2r-1}\right]}},~&u\in[0,1-p),\\
\mu,~&u\in[1-p,p),\\
\mu+\sigma\frac{1-p+2\tau[-r (1-u)^{r-1}+(1-p)^{r-1}]}{\sqrt{2\left[(1-p)^{3}+\frac{4\tau^{2}(r-1)^{2}(1-p)^{2r-1}}{2r-1}\right]}},~&u\in[p,1],
\end{cases}
\end{align*}
when $p<\frac{1}{2}$,
\begin{align*}
   &\sup_{X\in V_{S} (\mu,\sigma)}\mathrm{EGS}_{r,p}^{\tau} (X)= \mu +\frac{\sigma}{\sqrt{2}}\\
   &\small\cdot\sqrt{\frac{p}{(1-p)^{2}}+4\tau p(1-p)^{r-4}-\frac{4\tau p^{r}}{(1-p)^{3}}+4\tau^{2}p(1-p)^{2r-6}-8\tau^{2}p^{r}(1-p)^{r-5}+\frac{4\tau^{2}r^{2}(1-p)^{2r-5}}{2r-1}-\frac{4\tau^{2}r^{2}[\mathbf{B}_{1-p}(r,r)-\mathbf{B}_{p}(r,r)]}{(1-p)^{4}}},
\end{align*}
which can be obtained by the
worst-case distribution of r.v. $X_{\ast}$ characterized by
\begin{align*}
F_{X_{\ast}}^{-1}(u)=
\begin{cases}
\mu-\sigma\frac{1-p+2\tau[-r u^{r-1}+(1-p)^{r-1}]}{\sqrt{2\left[{p}{(1-p)^{2}}+4\tau p(1-p)^{r}-{4\tau p^{r}}{(1-p)}+4\tau^{2}p(1-p)^{2r-2}-8\tau^{2}p^{r}(1-p)^{r-1}+\frac{4\tau^{2}r^{2}(1-p)^{2r-1}}{2r-1}-{4\tau^{2}r^{2}[\mathbf{B}_{1-p}(r,r)-\mathbf{B}_{p}(r,r)]}\right]}},~&u\in[0,p),\\
\mu+\sigma\frac{\sqrt{2}\tau r[u^{r-1}-(1-u)^{r-1}]}{\sqrt{{p}{(1-p)^{2}}+4\tau p(1-p)^{r}-{4\tau p^{r}}{(1-p)}+4\tau^{2}p(1-p)^{2r-2}-8\tau^{2}p^{r}(1-p)^{r-1}+\frac{4\tau^{2}r^{2}(1-p)^{2r-1}}{2r-1}-{4\tau^{2}r^{2}[\mathbf{B}_{1-p}(r,r)-\mathbf{B}_{p}(r,r)]}}},~&u\in[p,1-p),\\
\mu+\sigma\frac{1-p+2\tau[-r (1-u)^{r-1}+(1-p)^{r-1}]}{\sqrt{2\left[{p}{(1-p)^{2}}+4\tau p(1-p)^{r}-{4\tau p^{r}}{(1-p)}+4\tau^{2}p(1-p)^{2r-2}-8\tau^{2}p^{r}(1-p)^{r-1}+\frac{4\tau^{2}r^{2}(1-p)^{2r-1}}{2r-1}-{4\tau^{2}r^{2}[\mathbf{B}_{1-p}(r,r)-\mathbf{B}_{p}(r,r)]}\right]}},~&u\in[1-p,1].
\end{cases}
\end{align*}
(iii) If $X\in V_{SU} (\mu,\sigma)$, by Theorem \ref{th.3}, then
  \begin{align*}
   &\sup_{X\in V_{SU} (\mu,\sigma)}\mathrm{EGS}_{r,p}^{\tau} (X)= \mu +\frac{2r(r-1)\tau\sigma}{(1-p)^{2}}\\
   &\cdot\left[\frac{2}{3}\int_{0}^{\frac{1}{3}}\sqrt{u}(1-u)^{r-2}\mathbb{I}_{[p,1]}(u)\mathrm{d}u+\sqrt{3}\int_{\frac{1}{3}}^{\frac{2}{3}}(1-u)^{r-1}u\mathbb{I}_{[p,1]}(u)\mathrm{d}u+\frac{2}{3}\int_{\frac{2}{3}}^{1}(1-u)^{r-3/2}\mathbb{I}_{[p,1]}(u)\mathrm{d}u\right],~\tau\neq0.
\end{align*}
\end{example}

Letting $r=2$ in Example \ref{ex.10}, $\mathrm{EGS}_{r,p}^{\tau} (X)$ will reduce to the Gini shortfall (GS) of a random variable $X$, with $p\in(0,1)$ and the loading
parameter $\tau \geq 0$, denoted by $\mathrm{GS}_{p}^{\tau} (X)$ (see Furman et al. (2017)), defined by
 \begin{align*}
\mathrm{GS}_{p}^{\tau} (X) = \mathrm{ES}_{p}(X) + \tau \mathrm{TGini}_{p}(X),
\end{align*}
where $
 \mathrm{TGini}_{p}(X)=-\int_{x_{p}}^{+\infty}2\left[\left(\frac{\overline{F}_{X}(x)}{1-p}\right)^{2}-\frac{\overline{F}_{X}(x)}{1-p}\right]\mathrm{d}x
 $
 is tail-Gini functional.\\
(i) If $X\in V_{U} (\mu,\sigma)$, then
\begin{align*}
   \sup_{X\in V_{U} (\mu,\sigma)}\mathrm{GS}_{p}^{\tau} (X)=\mu +\frac{4\tau\sigma}{3(1-p)^{2}}\left[\int_{0}^{1/2}\sqrt{u(8-9u)}\mathbb{I}_{[p,1]}(u)\mathrm{d}u+\int_{1/2}^{1}\sqrt{(9u-1)(1-u)}\mathbb{I}_{[p,1]}(u)\mathrm{d}u\right],~\tau\neq0.
\end{align*}
(ii) If $X\in V_{S} (\mu,\sigma)$, then\\
when $p\geq\frac{1}{2}$,
\begin{align*}
   \sup_{X\in V_{S} (\mu,\sigma)}\mathrm{GS}_{p}^{\tau} (X)= \mu +{\sigma}\sqrt{\frac{1}{2}\left[\frac{1}{1-p}+\frac{4\tau^{2}}{3(1-p)}\right]},
\end{align*}
which can be obtained by the
worst-case distribution of r.v. $X_{\ast}$ characterized by
\begin{align*}
F_{X_{\ast}}^{-1}(u)=
\begin{cases}
\mu-\sigma\frac{1-p+2\tau[-2 u+(1-p)]}{\sqrt{2\left[(1-p)^{3}+\frac{4\tau^{2}(1-p)^{3}}{3}\right]}},~&u\in[0,1-p),\\
\mu,~&u\in[1-p,p),\\
\mu+\sigma\frac{1-p+2\tau[-r (1-u)+(1-p)]}{\sqrt{2\left[(1-p)^{3}+\frac{4\tau^{2}(1-p)^{3}}{3}\right]}},~&u\in[p,1],
\end{cases}
\end{align*}
when $p<\frac{1}{2}$,
\begin{align*}
   &\sup_{X\in V_{S} (\mu,\sigma)}\mathrm{GS}_{p}^{\tau} (X)= \mu +\frac{\sigma}{\sqrt{2}}
   \sqrt{\frac{p(1+2\tau)^{2}}{(1-p)^{2}}-\frac{4\tau p^{2}(1+2\tau)}{(1-p)^{3}}+\frac{16\tau^{2}}{3(1-p)}-\frac{16\tau^{2}[\mathbf{B}_{1-p}(2,2)-\mathbf{B}_{p}(2,2)]}{(1-p)^{4}}},
\end{align*}
which can be obtained by the
worst-case distribution of r.v. $X_{\ast}$ characterized by
\begin{align*}
F_{X_{\ast}}^{-1}(u)=
\begin{cases}
\mu-\sigma\frac{1-p+2\tau[-2 u+(1-p)]}{\sqrt{2\left[{p}{(1-p)^{2}}(1+2\tau)^{2}-{4\tau p^{2}}{(1-p)}(1+2\tau)+\frac{16\tau^{2}(1-p)^{3}}{3}-{16\tau^{2}[\mathbf{B}_{1-p}(2,2)-\mathbf{B}_{p}(2,2)]}\right]}},~&u\in[0,p),\\
\mu+\sigma\frac{2\sqrt{2}\tau [2u-1]}{\sqrt{{p}{(1-p)^{2}}(1+2\tau)^{2}-{4\tau p^{2}}{(1-p)}(1+2\tau)+\frac{16\tau^{2}(1-p)^{3}}{3}-{16\tau^{2}[\mathbf{B}_{1-p}(2,2)-\mathbf{B}_{p}(2,2)]}}},~&u\in[p,1-p),\\
\mu+\sigma\frac{1-p+2\tau[-2 (1-u)+(1-p)]}{\sqrt{2\left[{p}{(1-p)^{2}}(1+2\tau)^{2}-{4\tau p^{2}}{(1-p)}(1+2\tau)+\frac{16\tau^{2}(1-p)^{3}}{3}-{16\tau^{2}[\mathbf{B}_{1-p}(2,2)-\mathbf{B}_{p}(2,2)]}\right]}},~&u\in[1-p,1].
\end{cases}
\end{align*}(iii) If $X\in V_{SU} (\mu,\sigma)$, then
\begin{align*}
&\sup_{X\in V_{SU} (\mu,\sigma)}\mathrm{GS}_{p}^{\tau} (X)\\
&= \mu +\frac{4\tau\sigma}{(1-p)^{2}}
   \left[\frac{2}{3}\int_{0}^{\frac{1}{3}}\sqrt{u}\mathbb{I}_{[p,1]}(u)\mathrm{d}u+\sqrt{3}\int_{\frac{1}{3}}^{\frac{2}{3}}(1-u)u\mathbb{I}_{[p,1]}(u)\mathrm{d}u+\frac{2}{3}\int_{\frac{2}{3}}^{1}(1-u)^{1/2}\mathbb{I}_{[p,1]}(u)\mathrm{d}u\right],~\tau\neq0.
   \end{align*}

\begin{remark}\label{re.6}
Note that when $\tau=0$ in Example \ref{ex.10}, above shortfall will reduce to $\mathrm{ES}_{p}(X)$. In this case, $\hat{g}'(u)=\frac{1}{1-p}\mathbb{I}_{[p,1]}(u)$,\\
 (i) If $X\in V_{U} (\mu,\sigma)$,  then

\begin{align}\label{yy5}
   \sup_{X\in V_{U} (\mu,\sigma)}\mathrm{ES}
_{p}(X)=
 \begin{cases}
 \mu+\frac{\sqrt{p(8-9p)}}{3(1-p)}\sigma,p\in(0,\frac{1}{2}],\\
  \mu+\frac{1}{3}\sqrt{\frac{9p-1}{1-p}}\sigma,p\in(\frac{1}{2},1).\\
\end{cases}
\end{align}
 (ii) If $X\in V_{S} (\mu,\sigma)$,  then\\
 when $p\geq \frac{1}{2}$,
 \begin{align*}
   \sup_{X\in V_{S} (\mu,\sigma)}\mathrm{ES}
_{p}(X)= \mu +\sigma\sqrt{\frac{1}{2(1-p)}},
\end{align*}
 which can be obtained by the
worst-case distribution of r.v. $X_{\ast}$ characterized by
\begin{align*}
F_{X_{\ast}}^{-1+}(u)=
\begin{cases}
\mu-\sigma\frac{1}{\sqrt{2(1-p)}},~&u\in[0,1-p),\\
\mu,~&u\in[1-p,p),\\
\mu+\sigma\frac{1}{\sqrt{2(1-p)}},~&u\in[p,1],
\end{cases}
\end{align*}
when $p< \frac{1}{2}$,
\begin{align*}
  & \sup_{X\in V_{S} (\mu,\sigma)}\mathrm{ES}
_{p}(X)
=\mu +\sigma\sqrt{\frac{p}{2(1-p)^{2}}},
\end{align*}
 which can be obtained by the
worst-case distribution of r.v. $X_{\ast}$ characterized by
\begin{align*}
F_{X_{\ast}}^{-1+}(u)=
\begin{cases}
\mu-\sigma\frac{1}{\sqrt{{2p}}},~&u\in[0,p),\\
\mu,~&u\in[p,1-p),\\
\mu+\sigma\frac{1}{\sqrt{{2p}}},~&u\in[1-p,1].
\end{cases}
\end{align*}
 (iii) If $X\in V_{SU} (\mu,\sigma)$,  then
\begin{align}\label{yy6}
\sup_{X\in V_{SU} (\mu,\sigma)}\mathrm{ES}
_{p}(X)=
 \begin{cases}
 \mu +\sigma\frac{2\sqrt{p}}{3(1-p)},~p\in(0,\frac{1}{3}),\\
 \mu +\sigma\sqrt{3}p,~p\in[\frac{1}{3},\frac{2}{3}),\\
\mu +\sigma\frac{2}{3\sqrt{1-p}},~p\in[\frac{2}{3},1).
\end{cases}
\end{align}
These results are a generalization of Lemmm 2.6 in Li et al. (2018), which requires $p\in(\frac{5}{6},1)$. In addition, Eq. (\ref{yy5}) is different from Corollary 4 of Bernard et al. (2020), where the set in Bernard et al. (2020) requires $\mathrm{Var}(X)\leq\sigma^{2}$ instead of $\mathrm{Var}(X)=\sigma^{2}$. Eq. (\ref{yy6}) is clearly different from Theorem 4.1 ($\beta\rightarrow1$) of Bernard et al. (2025), where the set in Bernard et al. (2025) also requires $\mathrm{Var}(X)\leq\sigma^{2}$ instead of $\mathrm{Var}(X)=\sigma^{2}$.
\end{remark}

\section{Applications to other risk measures}
In this section, we give applications to several risk measures, including $n$th-order expected shortfall, dual power
principle and proportional hazard
principle. Only the $n$th-order expected shortfall is given here, and the rest is included in the Supplementary Material.

\begin{example}\label{ex.13} Let $\hat{g}(u)=\left(\frac{u-p}{1-p}\right)^{n}\mathbb{I}_{[p,1]}(u)$, $u\in[0,1]$, $n\in\mathbb{N}$, and $g(1)=\hat{g}(1)=1$ in Lemma \ref{le.1}, it is called as
the $n$th-order expected shortfall of a random variable $X\in L^{1}$ at a level $p \in [0, 1)$,
  denoted by $\mathrm{ES}_{n,p}(X)$ (see, e.g., Fuchs et al. (2017), Barczy et al. (2023), Zou and Hu (2024)), defined as
  \begin{align*}
\mathrm{ES}_{n,p}(X)=\frac{n}{1-p}\int_{p}^{1}\left(\frac{u-p}{1-p}\right)^{n-1}F_{X}^{-1}(u)\mathrm{d}u,~n\in\mathbb{N}.
\end{align*}
 (i) If $X\in V_{U} (\mu,\sigma)$, by Theorem \ref{th.1}, then
 \begin{align*}
   &\sup_{X\in V_{U} (\mu,\sigma)}\mathrm{ES}_{n,p}(X)= \\ &\mu+
   \frac{n(n-1)\sigma}{3(1-p)^{n}}\left[\int_{0}^{1/2}\sqrt{u(8-9u)}(u-p)^{n-2}\mathbb{I}_{[p,1]}(u)\mathrm{d}u+\int_{1/2}^{1}\sqrt{(9u-1)(1-u)}(u-p)^{n-2}\mathbb{I}_{[p,1]}(u)\mathrm{d}u\right],~n\geq2.
\end{align*}
 (ii) If $X\in V_{S} (\mu,\sigma)$, by Theorem \ref{th.2}, then\\
 when $p\geq\frac{1}{2}$,
\begin{align*}
\sup_{X\in V_{S}(\mu,\sigma)}\mathrm{ES}_{n,p}(X)=\mu+\frac{\sigma}{2}\sqrt{\frac{2n^{2}}{(2n-1)(1-p)}},
\end{align*}
which can be obtained by the worst-case
distribution of r.v. $X_{\ast}$ characterized by
\begin{align*}
F_{X_{\ast}}^{-1+}(u)=
\begin{cases}
\mu-\sigma\frac{\frac{n}{1-p}\left(\frac{1-u-p}{1-p}\right)^{n-1}}{\sqrt{\frac{2n^{2}}{(2n-1)(1-p)}}},~&u\in[0,1-p),\\
\mu,~&u\in[1-p,p),\\
\mu+\sigma\frac{\frac{n}{1-p}\left(\frac{u-p}{1-p}\right)^{n-1}}{\sqrt{\frac{2n^{2}}{(2n-1)(1-p)}}},~&u\in[p,1],
\end{cases}
\end{align*}
when $p<\frac{1}{2}$,
\begin{align*}
\sup_{X\in V_{S}(\mu,\sigma)}\mathrm{ES}_{n,p}(X)=\mu+\frac{\sigma}{2}\sqrt{\frac{2n^{2}}{(1-p)^{2n}}\left[\frac{(1-p)^{2n-1}}{2n-1}-\kappa\right]},
\end{align*}
which can be obtained by the worst-case
distribution of r.v. $X_{\ast}$ characterized by
\begin{align*}
F_{X_{\ast}}^{-1+}(u)=
\begin{cases}
\mu-\sigma\frac{\frac{n}{1-p}\left(\frac{1-u-p}{1-p}\right)^{n-1}}{\sqrt{\frac{2n^{2}}{(1-p)^{2n}}\left[\frac{(1-p)^{2n-1}}{2n-1}-\kappa\right]}},~&u\in[0,p),\\
\mu+\sigma\frac{\frac{n}{1-p}\left(\frac{u-p}{1-p}\right)^{n-1}-\frac{n}{1-p}\left(\frac{1-u-p}{1-p}\right)^{n-1}}{\sqrt{\frac{2n^{2}}{(1-p)^{2n}}\left[\frac{(1-p)^{2n-1}}{2n-1}-\kappa\right]}},~&u\in[p,1-p),\\
\mu+\sigma\frac{\frac{n}{1-p}\left(\frac{u-p}{1-p}\right)^{n-1}}{\sqrt{\frac{2n^{2}}{(1-p)^{2n}}\left[\frac{(1-p)^{2n-1}}{2n-1}-\kappa\right]}},~&u\in[1-p,1],
\end{cases}
\end{align*}
where $\kappa=\int_{p}^{1-p}(u-p)^{n-1}(1-u-p)^{n-1}\mathrm{d}u.$\\
(iii) If $X\in V_{SU} (\mu,\sigma)$, by Theorem \ref{th.3}, then
\begin{align*}
   &\sup_{X\in V_{SU} (\mu,\sigma)}\mathrm{ES}_{n,p}(X)= \mu+\frac{n(n-1)\sigma}{(1-p)^{n}}\\ &\cdot\left[\frac{2}{3}\int_{0}^{\frac{1}{3}}\sqrt{u}(u-p)^{n-2}\mathbb{I}_{[p,1]}(u)\mathrm{d}u+\sqrt{3}\int_{\frac{1}{3}}^{\frac{2}{3}}(1-u)u(u-p)^{n-2}\mathbb{I}_{[p,1]}(u)\mathrm{d}u+\frac{2}{3}\int_{\frac{2}{3}}^{1}\sqrt{1-u}(u-p)^{n-2}\mathbb{I}_{[p,1]}(u)\mathrm{d}u\right],~n\geq2.
\end{align*}
\end{example}
\begin{remark}
Note that  when $n=1$, $\mathrm{ES}_{n,p}(X)$ will reduce to $\mathrm{ES}_{p}(X)$. Then we obtain the same results as in Remark \ref{re.6}.
\end{remark}
\section{Numerical illustration}
In this section, we consider daily stock returns of three stock companies (Coca-Cola Company
(KO), Cincinnati Financial (CINF),  eBay Inc. (EBAY)), and denote by $X_{1}$, $X_{2}$ and $X_{3}$, respectively. The data are from the
workdays for the period 2024-05-20 to 2024-05-18, (For the data, see http://www.nasdaq.com/).  Their mean and variance can be computed, respectively written as $(\mu_{1}=0.052128514,\sigma_{1}^{2}=0.416460169)$, $(\mu_{2}=-0.009317269
,\sigma_{2}^{2}=3.604281185)$ and $(\mu_{3}=0.09911245,\sigma_{3}^{2}=0.900906979)$.  Their histograms of daily stock returns of three stock companies can be plotted in Fig. 5.

Now, we calculate and compare the sharp upper bounds of Gini shortfall ($GS_{p}^{\tau}$) for general, unimodal, symmetric and  symmetric unimodal distributions under partial information constraints. Let $\tau=0.25$, $r=2$, $p=0.80,~0.90,~0.95, ~0.98, ~0.99$ in Example 6.2 of Zuo and Yin (2025) and Example \ref{ex.10}, respectively. From Fig. 5, we can observe that the distributions of $X_{1}$, $X_{2}$ and $X_{3}$ almost all follow unimodel and symmetric distributions.  The sharp upper bounds of $GS_{p}^{\tau}$ of the same $X_{i}$, $i = 1,2,3$ for general, unimodal, symmetric and  symmetric unimodal distributions in $p\in\{0.80,~0.90,~0.95, ~0.98, ~0.99\}$ are presented in Tables 1-3, respectively.
\begin{table}[!htbp]
\centering Table 1: The sharp upper bounds of $GS_{p}^{\tau}(X_{1})$ for general, unimodal, symmetric and  symmetric unimodal distributions in $p\in\{0.80,~0.90,~0.95, ~0.98, ~0.99\}$.\\
\begin{tabular}{cccccccccccccccccccccccc}
  \hline
  p& 0.80 &0.90 &0.95&0.98&0.99\\
  \hline
  $\sup_{X_{1}\in V (\mu_{1},\sigma_{1})} GS_{p}^{\tau}(X_{1})$&1.408361      &2.075787&2.985871&4.757636& 6.737939  \\
$\sup_{X_{1}\in V_{U} (\mu_{1},\sigma_{1})} GS_{p}^{\tau}(X_{1})$& 0.8951964     &1.290611&1.83519&2.900856& 4.09461  \\
$\sup_{X_{1}\in V_{S} (\mu_{1},\sigma_{1})} GS_{p}^{\tau}(X_{1})$&  1.114161    &1.554069&2.176193& 3.41057&  4.801682\\
$\sup_{X_{1}\in V_{SU} (\mu_{1},\sigma_{1})} GS_{p}^{\tau}(X_{1})$& 0.6934698     &0.9591221&1.334811&2.080228&2.920294   \\
  \hline
\end{tabular}
\end{table}
\begin{table}[!htbp]
\centering Table 2: The sharp upper bounds of $GS_{p}^{\tau}(X_{2})$ for general, unimodal, symmetric and  symmetric unimodal distributions in $p\in\{0.80,~0.90,~0.95, ~0.98, ~0.99\}$.\\
\begin{tabular}{cccccccccccccccccccccccc}
  \hline
  p& 0.80 &0.90 &0.95&0.98&0.99\\
  \hline
  $\sup_{X_{2}\in V (\mu_{2},\sigma_{2})} GS_{p}^{\tau}(X_{2})$&  3.980533    &5.944009&8.621353&13.83365&19.65943   \\
$\sup_{X_{2}\in V_{U} (\mu_{2},\sigma_{2})} GS_{p}^{\tau}(X_{2})$&2.470874      &3.634128&5.236208&8.37125&11.88311   \\
$\sup_{X_{2}\in V_{S} (\mu_{2},\sigma_{2})} GS_{p}^{\tau}(X_{2})$& 3.115038     &4.409188&6.239393&9.870761& 13.96322  \\
$\sup_{X_{2}\in V_{SU} (\mu_{2},\sigma_{2})} GS_{p}^{\tau}(X_{2})$&  1.877422    &2.658935&3.764161&5.957075& 8.428436  \\
  \hline
\end{tabular}
\end{table}
\begin{table}[!htbp]
\centering Table 3: The sharp upper bounds of $GS_{p}^{\tau}(X_{3})$ for general, unimodal, symmetric and  symmetric unimodal distributions in $p\in\{0.80,~0.90,~0.95, ~0.98, ~0.99\}$.\\
\begin{tabular}{cccccccccccccccccccccccc}
  \hline
  p& 0.80 &0.90 &0.95&0.98&0.99\\
  \hline
  $\sup_{X_{3}\in V (\mu_{3},\sigma_{3})} GS_{p}^{\tau}(X_{3})$& 2.093857     & 3.075506&4.414057&7.019967&9.932593   \\
$\sup_{X_{3}\in V_{U} (\mu_{3},\sigma_{3})} GS_{p}^{\tau}(X_{3})$& 1.339096     &1.92067&2.721637&4.289016& 6.044789  \\
$\sup_{X_{3}\in V_{S} (\mu_{3},\sigma_{3})} GS_{p}^{\tau}(X_{3})$&   1.661148    &2.308165&3.223184&5.038704& 7.08475  \\
$\sup_{X_{3}\in V_{SU} (\mu_{3},\sigma_{3})} GS_{p}^{\tau}(X_{3})$&  1.042396    &1.433117&1.98568&3.082038& 4.317607  \\
  \hline
\end{tabular}
\end{table}

Tables 1-3 display sharp upper bounds of the daily stock returns of three stock companies for different distributions in $p\in\{0.80,~0.90,~0.95, ~0.98, ~0.99\}$, respectively. We can observe that all sharp upper bounds of $X_{i}$, $i=1,2,3$,  are increasing in $p$. For same $p$, $\sup_{X_{i}\in V (\mu_{i},\sigma_{i})} GS_{p}^{\tau}(X_{i})>\sup_{X_{i}\in V_{S} (\mu_{i},\sigma_{i})} GS_{p}^{\tau}(X_{i})>\sup_{X_{i}\in V_{U} (\mu_{i},\sigma_{i})} GS_{p}^{\tau}(X_{i})>\sup_{X_{i}\in V_{SU} (\mu_{i},\sigma_{i})} GS_{p}^{\tau}(X_{i})$, $i=1,2,3$. It reveals that the unimodal information has more influence on the sharp upper bounds than the symmetric information. Moreover, for same $p$ and distribution (shape), sharp upper bound of $X_{2}$ is largest among the three stock and sharp upper bound of $X_{1}$ is  least among the three stock. It indicates that the chosen of distribution (shape) seems to have not impacts on the size relationship of the sharp upper bounds of Gini shortfall of three stock companies.

 \section{Concluding remarks\label{sec:5}}
This paper has derived the lower and upper bounds of worst-case distortion riskmetrics and weighted entropy for unimodal, symmetric, and symmetric unimodal distributions with known mean and variance, which are generalizations of results of Bernard et al. (2020) and  Zhao et al. (2024). These results  are also applied to many general risk measures and variability measures. Moreover, future research will include (i) the optimal reinsurance problems in distortion riskmetrics for unimodal and symmetric unimodal distributions with partial information, and (ii) the sharp upper bounds of distortion riskmetrics and weighted entropy for multimodal distributions with partial information. It is also interesting to study these results for other distortion measures, such as the family of parameterized Wang premia (see, e.g., Fiori and Rosazza Gianin, (2023)).
\section*{ CRediT authorship contribution statement}
\noindent$\mathbf{Baishuai~ Zuo:}$ Investigation, Methodology, Writing-original draft, Writing-review \& editing, Software.\\
$\mathbf{ Chuancun~ Yin:}$ Conceptualization, Investigation, Methodology, Supervision, Validation, Writing-original draft, Writing-review \& editing.
\section*{Acknowledgments}

\noindent  The authors thank Professor Taizhong Hu for his helpful discussions and bringing up relevant references. The research was supported by the National Natural Science Foundation of China (No. 12071251).
\section*{Declaration of competing interest}
\noindent There is no competing interest.

\medskip


\section*{References}


\begin{thebibliography}{99}
\bibitem{Abramowitz1965}Abramowitz, M.,  Stegun, I.A., 1965. Handbook of Mathematical Functions with Formulas and Mathematical Tables. Dover, New York.
\bibitem{Asadi} Asadi, M., Zohrevand, Y., 2007. On the dynamic cumulative residual entropy. Journal of Statistical Planning and Inference 137, 1931-1941.
\bibitem{Barczy2023} Barczy, M.,  Ned$\acute{e}$nyi, F. K.,  S$\ddot{u}$t\H{o}, L., 2023. Probability equivalent level of Value at Risk and higher-order Expected
Shortfalls. Insurance: Mathematics and Economics 108, 107-128.

\bibitem{Berkhouch2018}Berkhouch M., Lakhnatia G., Righi M.B., 2018. Extended Gini-type measures of risk and variability. Applied Mathematical Finance 25 (3), 295-314.

\bibitem{Bernard2020}Bernard, C., Kazzi, R., Vanduffel, S., 2020. Range Value-at-Risk
bounds for unimodal distributions under partial information. Insurance:
Mathematics and Economics 94, 9-24.

\bibitem{Bernard2025} Bernard, C., Kazzi, R., Vanduffel, S., 2025. Model uncertainty assessment for symmetric
and right-skewed distributions. Scandinavian Actuarial Journal 2025, 510-531.


\bibitem{Bernard2024}Bernard, C., Pesenti, S., Vanduffel, S., 2024. Robust distortion risk
measures. Mathematical Finance 34 (3), 774-818.



\bibitem{Boyd2004}Boyd, S.,  Vandenberghe, L., 2004. Convex Optimization.  Cambridge University Press, New York.

\bibitem{Cai2025a} Cai, J., Li, J.Y.M., Mao, T., 2025a. Distributionally robust optimization
 under distorted expectations. Operations Research 72 (3), 969-985.
\bibitem{Cai2025b}Cai, J., Liu, F., Yin, M., 2025b. Worst-case distortion risk measures of transformed losses with uncertain distributions lying in Wasserstein balls. ASTIN Bulletin. Published online 2025:1-25. doi:10.1017/asb.2025.10074.

\bibitem{Cali2017}Cal\`{\i}, C., Longobardi, M., Ahmadi, J., 2017. Some properties of cumulative Tsallis entropy. Physica A: Statistical Mechanics and its Applications 486, 1012-1021.
\bibitem{Chakraborty2023}Chakraborty, S., Pradhan, B., 2023. On weighted cumulative Tsallis residual and past entropy measures. Communications in Statistics - Simulation and Computation 52 (5), 2058-2072.
\bibitem{Chen2011}Chen, L., He, S., Zhang, S., 2011. Tight bounds for some risk measures,
 with applications to robust portfolio selection. Operations Research 59 (4), 847-865.

\bibitem{Denneberg1990}Denneberg, D., 1990. Premium calculation: Why standard deviation
should be replaced by absolute deviation. ASTIN Bulletin 20 (2), 181-190.


\bibitem{Denuit2005}Denuit, M., Dhaene, J., Goovaerts, M., Kaas, R., 2005. Actuarial Theory for Dependent Risks: Measures, Orders, and Models. Wiley, UK, Chichester.

\bibitem{Dhaene2012}Dhaene, J., Kukush, A., Linders, D., Tang, Q., 2012. Remarks on
quantiles and distortion risk measures. European Actuarial Journal 2, 319-328.


\bibitem{Di Crescenzo2021}Di Crescenzo, A., Kayal, S., Meoli, A., 2021. Fractional generalized cumulative entropy and its dynamic version. Communications in Nonlinear Science and Numerical Simulation 102, 105899.
\bibitem{Di Crescenzo2002}Di Crescenzo, A.,  Longobardi, M., 2002. Entropy-based measure of uncertainty in past lifetime distributions. Journal of Applied Probability 39 (2), 434-440.

\bibitem{ElGhaoui2003}El Ghaoui, L, Oks, M., Oustry, F., 2003. Worst-case value-at-risk and
robust portfolio optimization: A conic programming approach. Operations
 Research 51 (4), 543-556.
\bibitem{Finkelstein2015}Finkelstein, M., Vaupel, J. W., 2015. On random age and remaining lifetime for population of
items. Applied Stochastic Models in Business and Industry 31, 681-689.

\bibitem{Fiori2023} Fiori, A. M.,  Rosazza Gianin, E., 2023. Generalized PELVE and applications to risk measures. European Actuarial Journal 13, 307-339.

\bibitem{Fuchs2017}Fuchs, S., Schlotter, R., Schmidt, K., 2017. A review and some complements on quantile risk measures and their domain. Risks 5 (4), 59.

\bibitem{Furman2017}Furman, E., Wang, R., Zitikis, R., 2017. Gini-type measures of risk and variability:
Gini shortfall, capital allocations, and heavy-tailed risks. Journal of Banking $\&$ Finance  83,
70-84.
\bibitem{Haberman2011}Haberman, S., Khalaf-Allahb, M., Verrall, R., 2011. Entropy, longevity, and the cost of annuities. Insurance: Mathematics and Economics 48, 197-204.
\bibitem{Hu2020}Hu T., Chen O., 2020. On a family of coherent measures of variability. Insurance: Mathematics and Economics 95, 173-182.
\bibitem{Khintchine1938}Khintchine, A. Y., 1938. On unimodal distributions. Izvestiya Nauchno-Issledovatel'skogo Instituta
Matematiki i Mekhaniki 2 (2), 1-7.

\bibitem{LiL2018} Li, L., Shao, H., Wang, R., Yang, J., 2018. Worst-case range value-at risk
 with partial information. SIAM Journal on Financial Mathematics 9 (1), 190-218.
\bibitem{LiJ2018}Li, J.Y.M., 2018. Closed-form solutions for worst-case law invariant risk measures with appliction to robust protfolio optimization. Operations Research 66 (6), 1533-1541.
 \bibitem{Liu2020}Liu, F., Cai, J., Lemieux, C., Wang, R., 2020. Convex risk functionals:
representation and applications. Insurance: Mathematics and Economics 90, 66-79.

\bibitem{Mirali2017a}Mirali, M., Baratpour, S., 2017. Some results on weighted cumulative entropy. Journal of the Iranian Statistical Society  16 (2), 21-32.

\bibitem{Misagh2011}Misagh, F., Panahi, Y., Yari, G.H., Shahi, R., 2011. Weighted cumulative entropy and its estimation. IEEE International Conference on Quality and Reliability (ICQR): Bangkok, Thailand, 477-480.
\bibitem{Misra2018} Misra, N., Naqvi, S., 2018. Some unified results on stochastic properties of residual lifetimes at random times. Brazilian Journal of Probability and Statistics 32 (2), 422-436.
\bibitem{Pesenti2024}Pesenti,
S.M., Wang, Q.,  Wang, R., 2024. Optimizing distortion riskmetrics with distributional uncertainty. Mathematical Programming. https://doi.org/10.1007/s10107-024-02128-6.
\bibitem{Psarrakos2013}Psarrakos, G., Navarro, J., 2013. Generalized cumulative residual entropy and record values. Metrika 76 (5), 623-640.
\bibitem{Psarrakos2019}Psarrakos, G., Sordo, M.A., 2019. On a family of risk measures based on proportional
hazards models and tail probabilities. Insurance: Mathematics and Economics 86, 232-240.

 \bibitem{Psarrakos2024}Psarrakos, G., Toomaj, A., Vliora, P., 2024.   A family of variability measures based on the cumulative residual entropy and distortion functions. Insurance: Mathematics and Economics 114, 212-222.



\bibitem{Rao2004}Rao, M., Chen, Y., Vemuri, B., Wang, F., 2004. Cumulative residual entropy: A new measure of information. IEEE Transactions on Information Theory  50 (6), 1220-1228.
\bibitem{Ruschendorf2024}R\"{u}schendorf, L., Vanduffel, S., Bernard, C., 2024. Model Risk Management: Risk Bounds under Uncertainty. Cambridge University Press, New York.
\bibitem{Shao2023}Shao, H., Zhang, Z.G., 2023. Distortion risk measure under parametric
ambiguity. European Journal of Operational Research 311 (3), 1159-1172.

\bibitem{Shao2024}Shao, H., Zhang, Z.G., 2024. Extreme-case distortion risk measures: a
unification and generalization of closed-form solutions. Mathematics of Operations Research 49 (4), 2341-2355.
\bibitem{Suhov2015}Suhov, Y., Yasaei Sekeh, S., 2015. Weighted cumulative entropies: An extention of CRE and CE. arXiv: 1507.07051v1.
\bibitem{Sun2022}Sun, H., Chen, Y., Hu, T., 2022. Statistical inference for tail-based cumulative residual entropy. Insurance: Mathematics and Economics 103, 66-95.
\bibitem{Wang2022a}Wang, Q., Wang, R., Wei, Y., 2020a. Distortion riskmetrics on general spaces. ASTIN Bulletin 50 (3), 827-851.

  \bibitem{Wang2020b}Wang, R., Wei, Y., Willmot, G.E., 2020b. Characterization, robustness, and aggregation of signed Choquet
integrals. Mathematics of Operations Research 45 (3), 993-1015.

\bibitem{Wang2000}Wang, S.S., 2000. A class of distortion operators for pricing financial and insurance risks. Journal of Risk and Insurance 67, 15-36.

\bibitem{Xiong2019}Xiong, H., Shang, P., Zhang Y., 2019. Fractional cumulative residual entropy. Communications in Nonlinear Science and Numerical Simulation 78, 104879.

\bibitem{Yaari1987} Yaari, M., 1987. The dual theory of choice under risk. Econometrica 55, 95-115.

\bibitem{Yin2023} Yin, X.,  Balakrishnan, N.,  Yin, C., 2023. Bounds for Gini's mean difference based on first four moments, with some applications.  Statistical Papers 64 (6), 2081-2100.

\bibitem{Zhao2024} Zhao, M., Balakrishnan, N., Yin, C., Shao, H., 2024. Best-case and worst-case distortion risk measures
with partial information constrants.

 \bibitem{Zhu2018} Zhu, W., Shao, H., 2018. Closed-form solutions for extreme-case distortion
 risk measures and applications to robust portfolio management.
SSRN. https://ssrn.com/abstract=3103458.

\bibitem{Zou2024} Zou, Z.,  Hu, T., 2024.
Adjusted higher-order expected shortfall. Insurance: Mathematics and Economics 115, 1-12.

    \bibitem{Zuo2023}Zuo, B., Yin, C., 2023. Covariance representations and coherent measures for some entropies. Entropy 25 (11), 1525.

\bibitem{Zuo2025}Zuo, B., Yin, C., 2025. Worst-case distortion riskmetrics and weighted entropy with partial information. European Journal of Operational Research 321 (2), 476-492.

\end{thebibliography}
\end{document}